\documentclass[letterpaper, 9 pt, conference]{ieeeconf}      

\IEEEoverridecommandlockouts                              

\usepackage{graphics} 
\usepackage{epsfig} 
\usepackage{times} 
\usepackage{amsmath} 
\usepackage{amssymb}  
\usepackage{subfigure}
\usepackage{url}
\usepackage{dsfont}
\newtheorem{theorem}{Theorem}

\newtheorem{algorithm}[theorem]{Algorithm}

\usepackage{cite}

\usepackage{color}
\usepackage{algorithm}

\newtheorem{definition}[theorem]{Definition}

{ 

}

\newcommand{\bi}{\begin{itemize}}
\newcommand{\ei}{\end{itemize}}
\newcommand{\bd}{\begin{displaymath}}
\newcommand{\ed}{\end{displaymath}}
\newcommand{\be}{\begin{eqnarray*}}
\newcommand{\ee}{\end{eqnarray*}}

\usepackage[normalem]{ulem}

\title{\LARGE \bf
Data-driven Influence Based Clustering of Dynamical Systems}
\author{Subhrajit Sinha\\
\thanks{  S. Sinha is with Pacific Northwest National Laboratory, Richland, WA, USA - 99354\tt \small email : subhrajit.sinha@pnnl.gov
}
}

\begin{document}
\maketitle

\begin{abstract}
Community detection is a challenging and relevant problem in various disciplines of science and engineering like power systems, gene-regulatory networks, social networks, financial networks, astronomy etc. Furthermore, in many of these applications the underlying system is dynamical in nature and because of the complexity of the systems involved, deriving a mathematical model which can be used for clustering and community detection, is often impossible. Moreover, while clustering dynamical systems, it is imperative that the dynamical nature of the underlying system is taken into account. In this paper, we propose a novel approach for clustering dynamical systems purely from time-series data which inherently takes into account the dynamical evolution of the underlying system. In particular, we define a \emph{distance/similarity} measure between the states of the system which is a function of the influence that the states have on each other, and use the proposed measure for clustering of the dynamical system. For data-driven computation we leverage the Koopman operator framework which takes into account the nonlinearities (if present) of the underlying system, thus making the proposed framework applicable to a wide range of application areas. We illustrate the efficacy of the proposed approach by clustering three different dynamical systems, namely, a linear system, which acts like a proof of concept, the highly non-linear IEEE 39 bus transmission network and dynamic variables obtained from atmospheric data over the Amazon rain forest.
\end{abstract}

\section{Introduction}\label{section_Intro}

The scientific discipline of dynamical systems started with the revolutionary works of Sir Issac Newton \cite{newton_principia}. Since then it has matured into one of the most important branches of mathematics, with applications to a wide variety of fields. However, many practical systems of interest like power grids, biological networks, financial networks etc. are high dimensional systems with complicated topology. For such large systems, it often happens that the states of the system can be divided into subgroups such that the states belonging to a particular subgroup are \emph{similar} in some sense, whereas states belonging to different subgroups are not so \emph{similar}. This \emph{divide and conquer} policy for studying large systems is often helpful because the subgroups can be studied independent of one another, thus reducing the complexity of the problem. This approach of grouping \emph{similar things} together is quite general and can be applied to all systems which can be represented by a graph and is known as \emph{graph clustering} \cite{graph_theory_book, schaeffer2007graph}. The main idea behind clustering any general graph is to define a distance on the graph and then partitioning the graph such that the nodes belonging to one partition are \emph{close} to each other with respect to the distance, and nodes belonging to two different partitions are \emph{far} apart from each other with respect to the defined distance. See \cite{schaeffer2007graph} and the references therein.

Again, in the field of dynamical systems, until the later part of the last century, most of the studies of dynamical systems were theoretical; but with the advent and progress of computer technology, numerical and data-driven analysis of dynamical systems has become extremely popular. In fact, data-driven analysis of dynamical systems is often necessary and is the only way to analyze certain dynamical systems. This is because, many systems of interest, like power grids or biological networks, are extremely complicated and mathematical modelling of such systems from first principles is often impossible. Thus the only way to study such systems is using time-series data obtained from the system. But still one problem remains in the sense that even if the large dimensional system has been identified, analyzing the system may be difficult because of the high dimensionality of the underlying system and one efficient way out is clustering the dynamical states of the system so that one can analyze the individual clusters independently, thus reducing the difficulty of the problem. As an example, consider the problem of designing control strategies for a power grid. Implementing local controls is often both practical and cost-effective than implementing a global control. However, to implement local controls one has to make sure that the local control strategy that has been implemented to control a part of the network does not affect the rest of the network. Thus in this case it is necessary to cluster a power grid to identify buses or generators which are \emph{closely related} to each other and then implement local control strategies for each cluster. However, clustering of the states of a dynamical system should somehow take into account the dynamics of the system, because clustering, based only on the connectivity of states (static clustering) may generate irrelevant clusters.

Motivated by these (data-driven analysis and need for divide and conquer strategy), in this paper, we provide a purely data-driven framework for clustering the states of a dynamical system which explicitly takes into account the underlying dynamics of the system. In particular, we define a new notion of \emph{distance} between the states of a dynamical system and this distance, which is a measure of influence of the states on each other, can be used to define a directed weighted graph, which in turn can be used to group together (cluster) the \emph{similar} states of a general dynamical system. The notion of \emph{influence distance} is derived from the recently developed concept of information transfer between the states (subspaces) of a dynamical system \cite{sinha_IT_CDC2016,sinha_IT_ICC}, which has been shown to capture the intuitive notions of causality, namely, zero transfer, transfer asymmetry and information conservation \cite{sinha_IT_ICC}.  Apart from identifying the correct causal structure in a dynamical system, the information transfer measure also quantifies the influence of any state (subspace) on any other state (subspace) and the information measure has been used for identification of influential generators and states in a power network \cite{IT_influence_acc,sinha_cdc_2017_power} and also for characterization and classification of small-signal stability of a power grid \cite{sinha_power_journal_IEEEtran}.

For data-driven discovery, we use the Koopman operator framework \cite{mezic_koopmanism,mezic_spectral,sinha_equivariant} to approximate a dynamical system from time-series data and use the Koopman model to compute the information transfer measure \cite{sinha_IT_data_acc,sinha_IT_data_journal}. The information transfer measure is then used to define the similarity measure (distance) between the states of a dynamical system. This process generates a directed weighted graph (\emph{influence graph}), where the weights reflect the influence of the states on each other and existing clustering algorithms like spectral clustering, k-means clustering etc. can be used for grouping together \emph{similar states} of the dynamical system. Clustering of this directed graph partitions the system into clusters, such that, the states/subspaces belonging to the same cluster have high influence on each other, whereas states/subspaces belonging to different clusters have small influence on one another. 

The rest of the paper is organized as follows. In section \ref{section_IT}, we briefly discuss the notion of information transfer in a dynamical system, followed by a discussion on data-driven computation of information transfer in section \ref{section_data_IT}. In section \ref{section_Clustering} we define the influence distance and show via an example, how to define the weighted directed influence graph for a dynamical system. The simulation results are presented in section \ref{section_simulation}, where we present three different examples. Firstly, using a network of linear oscillators, we show why influence based clustering, which takes into account the underlying dynamics of a dynamical system, is more meaningful than clustering based only on the connectivity of a dynamical system. As the second example we use the proposed approach to cluster the generators of the IEEE 39 bus system by two different methods, namely, $k$-means clustering and hierarchical clustering. As the third example, we present preliminary results on clustering of dynamic variables present in the atmosphere over the Amazon rain forest. Finally, we conclude the paper in section \ref{section_Conclusions}.

\section{Information Transfer in Dynamical Systems}\label{section_IT}
Information transfer in a dynamical system \cite{sinha_IT_CDC2016,sinha_IT_ICC} characterizes and quantifies influence between the states (subspaces) of a dynamical system. In this section, we briefly review the concept of information transfer in a dynamical system. For details see \cite{sinha_IT_CDC2016,sinha_IT_ICC}. We consider a discrete time dynamical system
{\small
\begin{eqnarray}\label{system2d}\left.
\begin{array}{ccl}
x(t+1) = F_x(x(t),y(t))+\xi_x(t)\\
y(t+1) = F_y(x(t),y(t))+\xi_y(t)
\end{array}\right\}=F(z(t),\xi(t))
\end{eqnarray}
}
where $x\in\mathbb{R}^{|x|}$, $y\in\mathbb{R}^{|y|}$ (here $|\cdot|$ denotes the dimension of $\{\cdot\}$), $z=(x^\top,y^\top)^\top$, and  $F_x : \mathbb{R}^{|x|+|y|}\to\mathbb{R}^{|x|}$ and $F_y : \mathbb{R}^{|x|+|y|}\to\mathbb{R}^{|y|}$ are assumed to be continuously differentiable, $\xi(t)$ is assumed to be i.i.d. noise and let $\rho((\cdot)(t))$ denote the probability density of $(\cdot)$ at time $t$. With this, the information transfer from $x$ to $y$ is quantified in terms of the Shannon entropy of concerned variables. In particular, the the total entropy of $y$ is considered to be the sum of the information transferred from $x$ to $y$ and the entropy of $y$ when $x$ is forcefully not allowed to evolve and is held constant (frozen). To mimic the effect of \emph{freezing} of $x$, we consider the modified system
{\small
\begin{eqnarray}\label{system_xfreeze}\left.
\begin{array}{ccl}
x(t+1) &=& x(t)\\ 
y(t+1) &=& F_y(x(t),y(t)) + \xi_y(t) 
\end{array}\right\}=F_{\not x}(z(t),\xi(t)).
\end{eqnarray}
}
We denote by $\rho_{\not x}(y(t+1)|y(t))$ the probability density function of $y(t+1)$ conditioned on $y(t)$, with the dynamics in $x$ coordinate frozen in time going from time step $t$ to $t+1$ (Eq. (\ref{system_xfreeze})). With this, the information transfer from $x$ to $y$, as the dynamical system (\ref{system2d}) evolves from time step $t$ to $t+1$ is defined as:
\begin{definition}\label{IT_def}[Information transfer \cite{sinha_IT_CDC2016,sinha_IT_ICC}]  The information transfer from $x$ to $y$ for the dynamical system (\ref{system2d}), as the system evolves from time $t$ to time $t+1$ (denoted by $[T_{x\to y}]_t^{t+1}$), is given by following formula
\begin{eqnarray}\label{IT_formula}
[T_{x\to y}]_t^{t+1}=H(\rho(y(t+1)|y(t)))-H(\rho_{\not{x}}(y(t+1)|y(t)),
\end{eqnarray}
where $H(\rho(y(t)))=- \int_{\mathbb{R}^{|y|}} \rho(y(t))\log \rho(y(t))dy$ is the entropy of $y(t)$ and $H(\rho_{\not{x}}(y(t+1)|y(t))$ is the entropy of $y(t+1)$, conditioned on $y(t)$, where $x$ has been frozen. 
\end{definition}

The information transfer from $x$ to $y$ depicts how evolution of $x$ affects the evolution of $y$, that is, it gives a quantitative measurement of the influence of $x$ on $y$. With this, we say that $x$ causes $y$ or $x$ influences $y$ if and only if the information transfer from $x$ to $y$ is non-zero \cite{sinha_IT_CDC2016,sinha_IT_ICC}.

In this paper, we will consider steady state information transfer, which is defined as follows: 
\begin{definition}[Steady State Information Transfer]
The steady state information transfer from a state $x$ to state $y$ $(T_{x\to y}^{ss})$, for the dynamical system $z(t+1) = F(z(t),\xi(t))$ is defined as

{\small
\begin{eqnarray}\label{IT_def_ss}
T_{x\to y}^{ss} =\lim_{t\to \infty} [H(y(t+1)|y(t))-H_{\not{x}}(y(t+1)|y(t))],
\end{eqnarray}}
provided the limit exists and is finite.
\end{definition}

\subsection{Information transfer in linear dynamical systems}
For general nonlinear systems it is not possible to compute closed-form expression for information transfer. But for linear systems with additive Gaussian noise, it is possible to derive analytical expression for information transfer. Consider the following linear dynamical system 
\begin{eqnarray}
z(t+1)=Az(t)+\sigma \xi(t)\label{lti}
\end{eqnarray}
where $z(t) = [x^\top, y^\top]^\top= [x_1^\top, x_2^\top , y^\top]^\top \in \mathbb{R}^N$, $\xi(t)$ is vector valued Gaussian random variable with zero mean and unit variance and $\sigma>0$ is a constant. We assume that the initial conditions are Gaussian distributed with covariance $\Sigma(0)$. Then the information transfer from a state (subspace) $x_1$ to the state (subspace) $y$ is given by
\begin{eqnarray}\label{transferx1y}
[T_{x_1\to y}]_t^{t+1}=\frac{1}{2}\log \frac{|A_{yx}\Sigma^s_y(t)A_{yx}^\top +\sigma^2 I |}{|A_{yx_2}(\Sigma_y^{s})_{yx_2}(t)A_{yx_2}^\top+\sigma^2 I|}.
\end{eqnarray}
Here the system matrix $A$ and the covariance matrix $\Sigma$ is split as
\begin{eqnarray}
A=\begin{pmatrix}A_x&A_{xy}\\ A_{yx}&A_{y}\end{pmatrix}=\begin{pmatrix}A_{x_1}&A_{x_1x_2}& A_{x_1 y}\\A_{x_2x_1}&A_{x_2}& A_{x_2 y}\\ A_{y x_1}&A_{y x_2}& A_{y}\end{pmatrix}\label{splittingA}
\end{eqnarray}
and
\begin{eqnarray}
\Sigma=\begin{pmatrix}\Sigma_x&\Sigma_{xy}\\\Sigma_{xy}^\top& \Sigma_y\end{pmatrix}=\begin{pmatrix} \Sigma_{x_1}&\Sigma_{x_1x_2}&\Sigma_{x_1 y}\\\Sigma_{x_1x_2}^\top&\Sigma_{x_2}&\Sigma_{x_2 y}\\\Sigma_{x_1y}^\top&\Sigma_{x_2y}^\top&\Sigma_{y}\end{pmatrix}.
\label{sigma_dec}
\end{eqnarray}
Furthermore, $\Sigma^s_y(t)=\Sigma_x(t)-\Sigma_{xy}(t)\Sigma_y(t)^{-1}\Sigma_{xy}(t)^\top$ is the Schur complement of $\Sigma_{y}(t)$ in the matrix $\Sigma(t)$ and $ (\Sigma_y^s)_{yx_2}$ is the Schur complement of $\Sigma_{y}$ in the matrix 
\[\begin{pmatrix}\Sigma_{x_2}&\Sigma_{x_2y}\\\Sigma_{x_2 y}^\top&\Sigma_y\end{pmatrix},\]
where the covariance matrix $\Sigma(t)$ evolves according to the equation
\[A \Sigma(t-1)A^\top+\sigma^2 I=\Sigma(t).\]

For more details see \cite{sinha_IT_CDC2016,sinha_IT_ICC}.

\section{Data-driven Computation of Information Transfer}\label{section_data_IT}
In this section, we discuss the computation of information transfer from time-series data obtained from a dynamical system. For details, see \cite{sinha_IT_data_acc,sinha_IT_data_journal}.

Consider a data set ${\cal Z}  = [z_0,z_1,\ldots,z_M]$ obtained from a random dynamical system $z\mapsto T(z,\xi)$, where $z_i\in Z\subset \mathbb{R}^N$. The data-set $\{z_k\}$ can be viewed as sample path trajectory generated by random dynamical system and could be corrupted by either process or measurement noise or both. In the presence of noise Dynamic Mode Decomposition (DMD) \cite{schmid_DMD} or Extended Dynamic Mode Decomposition (EDMD) \cite{williams_EDMD,mezic_EDMD} algorithms often identify erroneous Koopman operator. The situation was salvaged in \cite{robust_dmd_acc,sinha_robust_DMD_journal}, where ideas from robust optimization were leveraged to propose an algorithm for computation of a Robust Koopman operator.

Let $\mathbf{\Psi}:Z\to \mathbb{R}^{K}$ be the set of observables which are used to lift the obtained data points from the state space $\mathbb{R}^N$ to a higher dimensional space $\mathbb{R}^K$, such that
\begin{equation}
\mathbf{\Psi}(z):=\begin{bmatrix}\psi_1(z) & \psi_2(z) & \cdots & \psi_K(z)\end{bmatrix}.\label{dic_function}
\end{equation}

With this, the robust Koopman operator $(\bf K)$ can be obtained as a solution to the following optimization problem \cite{robust_dmd_acc,sinha_robust_DMD_journal}

\begin{eqnarray}\label{rob_eqv}
\min\limits_{\bf K}\parallel {\bf K}{\bf Y}_p-{\bf Y}_f\parallel_F+\lambda \parallel {\bf K}\parallel_F
\end{eqnarray}
where
\begin{eqnarray}
\begin{aligned}
&{\bf Y}_p= {\bf \Psi}(X_p) = \begin{bmatrix}{\bf \Psi}(z_0) & {\bf \Psi}(z_1) & \cdots & {\bf \Psi}(z_{M-1})\end{bmatrix}\\
&{\bf Y}_f= {\bf \Psi}(X_f) = \begin{bmatrix}{\bf \Psi}(z_1) & {\bf \Psi}(z_2) & \cdots & {\bf \Psi}(z_{M})\end{bmatrix},
\end{aligned}
\end{eqnarray}
${\bf K}\in\mathbb{R}^{K\times K}$ is the robust Koopman operator and $\lambda$ is the regularization parameter which depends on the bounds of the process and measurement noise.

For computation of information transfer, we use the robust variant of the DMD algorithm, that is, we use ${\bf \Psi}(z) = z$. With this $\bar A=\bf K\in \mathbb{R}^{N\times N}$ is the estimated system dynamics obtained using optimization formulation (\ref{rob_eqv}). We further assume that the initial covariance matrix is $\bar \Sigma(0)$ so that the conditional entropy $H(y_{t+1}|y_t)$ for the non-freeze case is computed as   
\begin{eqnarray}\label{cond_entr}
H(y_{t+1}|y_t) = \frac{1}{2}\log |\bar A_{yx}\bar \Sigma_y^S(t)\bar A_{yx}^\top + \lambda I|,
\end{eqnarray}
where $|\cdot |$ denotes the determinant and the matrices $\bar A_{yx}$ and $\Sigma_y^S(t)$ are as defined in (\ref{splittingA}) and (\ref{sigma_dec}) respectively.

For computing the dynamics when $x$ is held frozen, one has to modify the obtained data so that it mimics the effect of holding $x$ constant at each time step. For simplicity, we describe the procedure for a two state system and the method generalizes easily for the $N$-dimensional case. Let the obtained time series data be given by

\begin{eqnarray}
\mathcal{Z}=\bigg[\begin{pmatrix}
x_0\\
y_0
\end{pmatrix},  \begin{pmatrix}
x_1\\
y_1
\end{pmatrix}, 
\cdots, \begin{pmatrix}
x_{M-1}\\
y_{M-1}
\end{pmatrix}\bigg]. \label{data_original}
\end{eqnarray}

From the obtained data-set we form a modified data-set as follows

\begin{eqnarray}
\mathcal{Z}_{\not{x}}=\bigg[\begin{pmatrix}
x_0\\
y_1
\end{pmatrix},  \begin{pmatrix}
x_1\\
y_2
\end{pmatrix}, 
\cdots, \begin{pmatrix}
x_{M-1}\\
y_{M}
\end{pmatrix}\bigg]. \label{data_frozen}
\end{eqnarray}

With this the system matrix for the frozen dynamics is computed as the solution of the optimization problem

\begin{eqnarray}\label{rob_eqv_froz}
\min\limits_{\bar{A}_{\not{x}}}\parallel {\bar{A}_{\not{x}}}{\cal Z}-{{\cal{Z}}_{\not{x}}}\parallel_F+\lambda \parallel {\bar{A}_{\not{x}}}\parallel_F
\end{eqnarray}
and the entropy $H_{\not{x}}(y_{t+1}|y_t)$ is obtained using equation (\ref{cond_entr}), with ${\bar A}$ and ${\bar \Sigma}$ replaced by the system matrix for the frozen dynamics and covariance matrix for the frozen dynamics respectively.


Finally the information transfer from $x$ to $y$ is computed using the formula
\[ [T_{x\to y}]_t^{t+1}=H(y_{t+1}|y_t)-H_{\not{x}}(y_{t+1}|y_t).\]

\begin{algorithm}[htp!]
\caption{Computation of Information Transfer from Time-series Data}
\begin{enumerate}
\item{From the original data set, compute the estimate of the system matrix $\bar A$ using the optimization formulation (\ref{rob_eqv}).}

\item{Assume $\bar \Sigma(0)$ and compute $\bar \Sigma(t)$ as 
\[\bar{\Sigma}(t) = \bar{A} \bar{\Sigma}(t-1)\bar{A}^\top+\sigma^2 I.\] Determine $\bar A_{yx}$ and $\bar \Sigma_y^S$ to calculate the conditional entropy $H(y_{t+1}|y_t)$ using (\ref{cond_entr}).}
\item{From the original data set form the modified data set mimicking the freezing of $x$, as given by Eq. (\ref{data_frozen}).}
\item{Compute the frozen dynamics using the optimization problem (\ref{rob_eqv_froz}).}
\item{Follow steps (1)-(2) with the frozen dynamics to compute the conditional entropy $H_{\not{x}}(y_{t+1}|y_t)$.}
\item{Compute the transfer $[T_{x\to y}]_t^{t+1}$ as $[T_{x\to y}]_t^{t+1} = H(y_{t+1}|y_t) - H_{\not{x}}(y_{t+1}|y_t)$.}
\end{enumerate}\label{algo_IT}
\end{algorithm}

\section{Influence Metric and Clustering}\label{section_Clustering}

In this section, we discuss the bare minimum basic concepts of graph theory \cite{graph_theory_book} that are required for this work and define a \emph{distance} between the states of a dynamical system, which we will call \emph{influence distance}.

A graph $G=(V,E)$ consists of a set of vertices (nodes) $V$ and a set of edges $E\subseteq V\times V$ which connect a pair of vertices. In this paper, we will consider directed graphs, where every edge $(i,j)\in E$ links node $i$ to node $j$. That is, a directed graph is a graph where the existence of an edge $(i,j)\in E$ implies there is a directed path from node $i$ to node $j$. Note that, in general $(i,j)\in E$ does not mean $(j,i)\in E$. 

Every such graph can be represented by a matrix, known as the adjacency matrix which is defined as follows:
\begin{definition}[Adjacency Matrix] \cite{graph_theory_book}
The adjacency matrix $A$ of a graph $G=(V,E)$ is a $|V|\times |V|$ matrix such that 
\begin{eqnarray}
A = \begin{cases}
w_{ij}, \textnormal{ if } (i,j)\in E\\
\infty, \textnormal{ otherwise.}
\end{cases}
\end{eqnarray}
\end{definition}
Here $w_{ij}$ is the weight of the edge $(i,j)$.

As mentioned earlier, the information transfer $T_{x\to y}$ gives the influence of $x$ on $y$. In particular, if $|T_{x\to y}|$ is large, then $x$ has a large influence on $y$ and if $|T_{x\to y}|$ is small then $x$ has very little influence on evolution of $y$. Using this notion of influence, we define the \emph{directed distance} from $x$ to $y$ as
\begin{definition}[Influence Distance] The influence distance from $x$ to $y$ is given by
\begin{eqnarray}
d(x,y) := \begin{cases}
\exp(-\frac{|T_{x\to y}|}{\beta}), \textnormal{ if } T_{x\to y}\neq 0\\
\infty, \textnormal{ if } T_{x\to y} = 0,
\end{cases}
\end{eqnarray}
where $\beta>0$ is a parameter analogous to temperature in the partition function of a Gibbs' distribution.
\end{definition}

Hence, if $x$ is transferring a lot of information to $y$ then $y$ is \emph{close} to $x$ and if $|T_{x\to y}|$ is small then $y$ is \emph{far} from $x$. For simulation purposes, if $T_{x\to y} = 0$, we set a large value for $d(x,y)$, typically of the order of $10^6$.

Now, given a dynamical system $z(t+1)=S(z(t)) + \xi(t)$, we define a weighted directed graph, such that each node of the graph correspond to a state of the system. Further, there is a directed edge from a node $x$ to node $y$ if and only if $T_{x\to y}\neq 0$, with the weight of the edge $(x,y)$ being $d(x,y)$ and in this paper, we use this influence distance as the similarity measure to cluster a dynamical system. 

For example, consider the 3-dimensional linear system given by
\begin{eqnarray}\label{3_state_sys}
\begin{aligned}
\begin{pmatrix}
x_1(t+1)\\
x_2(t+1)\\
x_3(t+1)
\end{pmatrix}=0.9\begin{pmatrix}
0 & 0 & 0\\
2 & 0 & 0.8\\
2 & 1 & 0
\end{pmatrix}\begin{pmatrix}
x_1(t)\\
x_2(t)\\
x_3(t)
\end{pmatrix}+\xi(t)
\end{aligned}
\end{eqnarray}
where $\xi(t)\in\mathbb{R}^3$ is an independent and identically distributed zero mean Gaussian noise of unit variance. 

\begin{figure}[htp!]
\centering
\subfigure[]{\includegraphics[scale=.225]{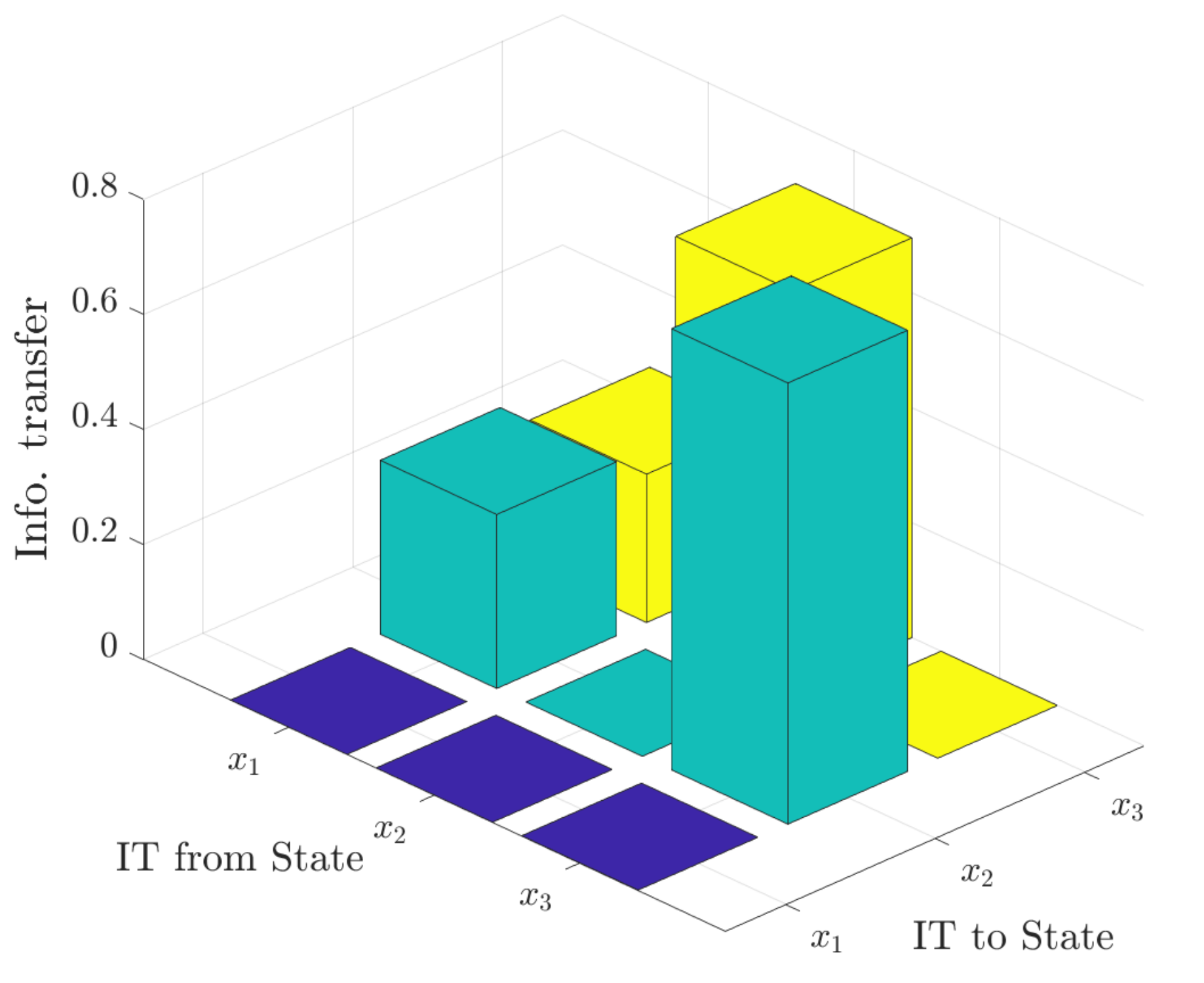}}
\subfigure[]{\includegraphics[scale=.32]{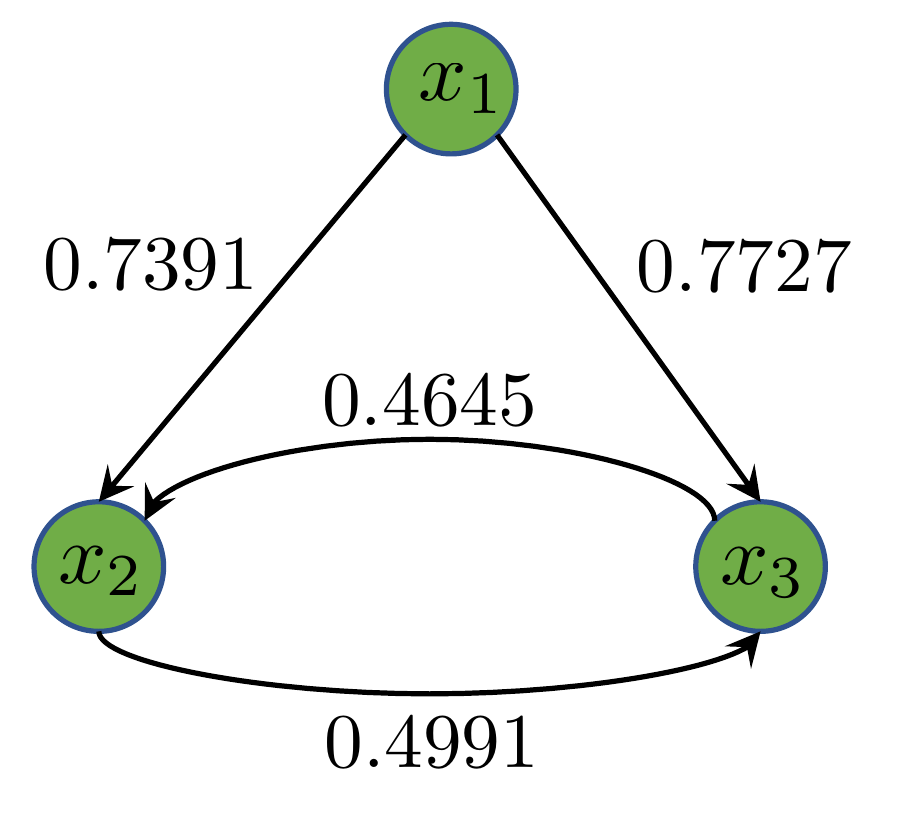}}
\caption{(a) Information transfer between the states. (b) Derived weighted directed graph of the system.}\label{fig_3_state_IT_graph}
\end{figure}

Then the steady state information transfer between the states is shown in Fig. \ref{fig_3_state_IT_graph}(a) and the corresponding weighted directed graph is shown in Fig. \ref{fig_3_state_IT_graph}(b). As an example, the information transfer from $x_1$ to $x_2$ is $T_{x_1\to x_2}=0.3023$ and hence $d(x_1,x_2) = exp(-0.3023)= 0.7391$. Hence, in the graph there is a directed edge from $x_1$ to $x_2$ with edge weight $0.7391$. The entire graph is obtained in a similar manner. Here we chose the parameter $\beta = 1$.

Once the weighted directed graph is obtained, the clustering of the nodes can be achieved by implementing the existing algorithms like $k$-means clustering \cite{kmeans} or hierarchical clustering \cite{hierarchical}. 

\section{Numerical Studies}\label{section_simulation}

\subsection{Clustering of a Network of Coupled Oscillators}

Consider a network of $N$ damped oscillators with equation of motion of each oscillator given by
\begin{eqnarray}\label{osc_sys}
\Ddot{\theta}_k = -{\cal L}_k\theta_k -d \dot{\theta}_k, \quad k = 1,2,\cdots , N,
\end{eqnarray}
where $\theta_k$ is the angular position of the $k^{th}$ oscillator, $N=12$ is the number of oscillators, ${\cal L}_k$ is the $k^{th}$ row of the Laplacian matrix $\cal L$ and $d=0.5$ is the damping co-efficient.
\eqref{osc_sys} can be re-written as:
\[
\frac{d}{dt}\begin{bmatrix}
          \theta_k \\
          \dot{\theta}_k 
         \end{bmatrix} = \begin{bmatrix}
0 & 1 \\
-{\cal L}_k & -d
\end{bmatrix} \begin{bmatrix}
          \theta_k \\
          \dot{\theta}_k 
         \end{bmatrix}.
\]


\begin{figure}[htp!]
\centering
\subfigure[]{\includegraphics[scale=.32]{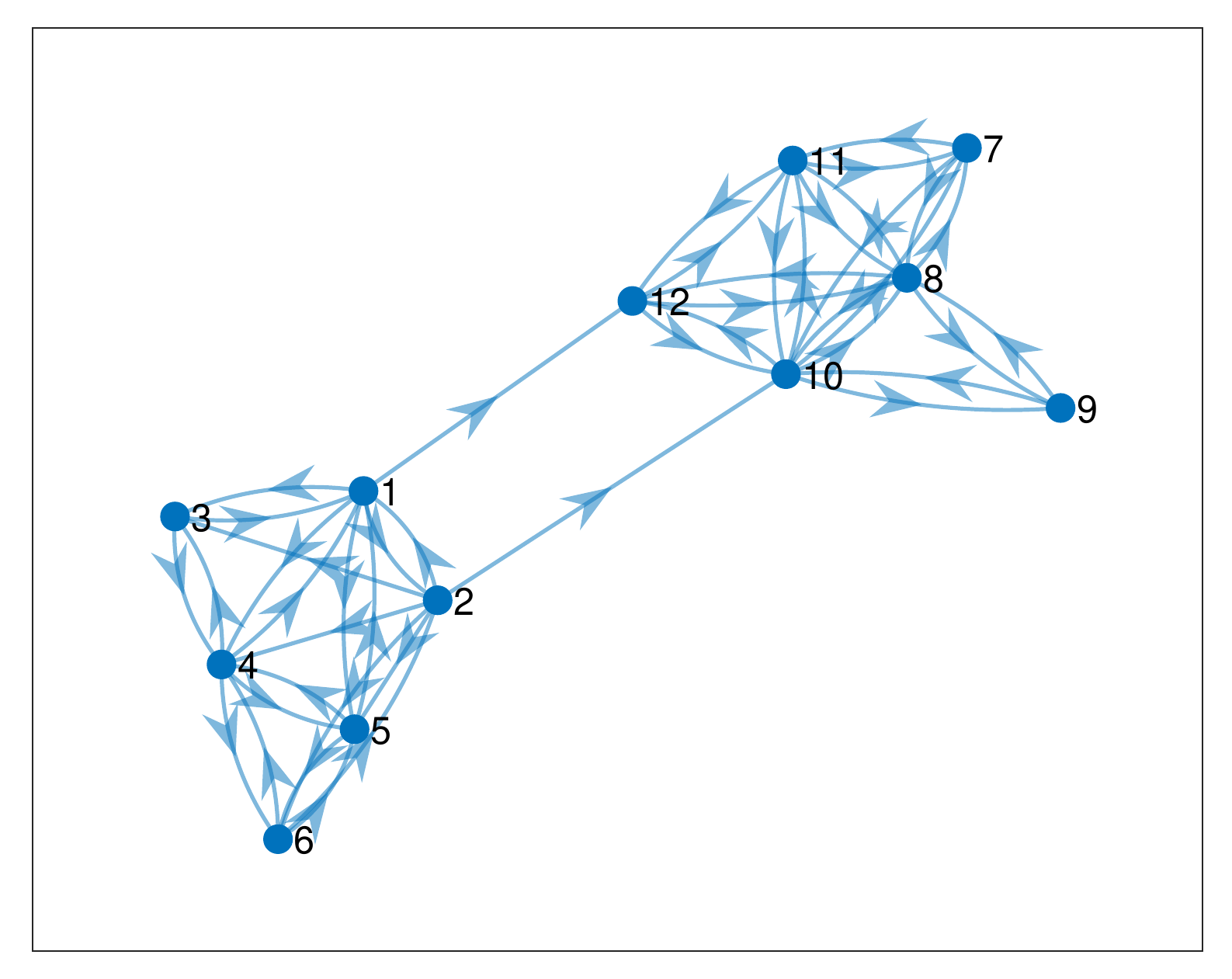}}
\subfigure[]{\includegraphics[scale=.3]{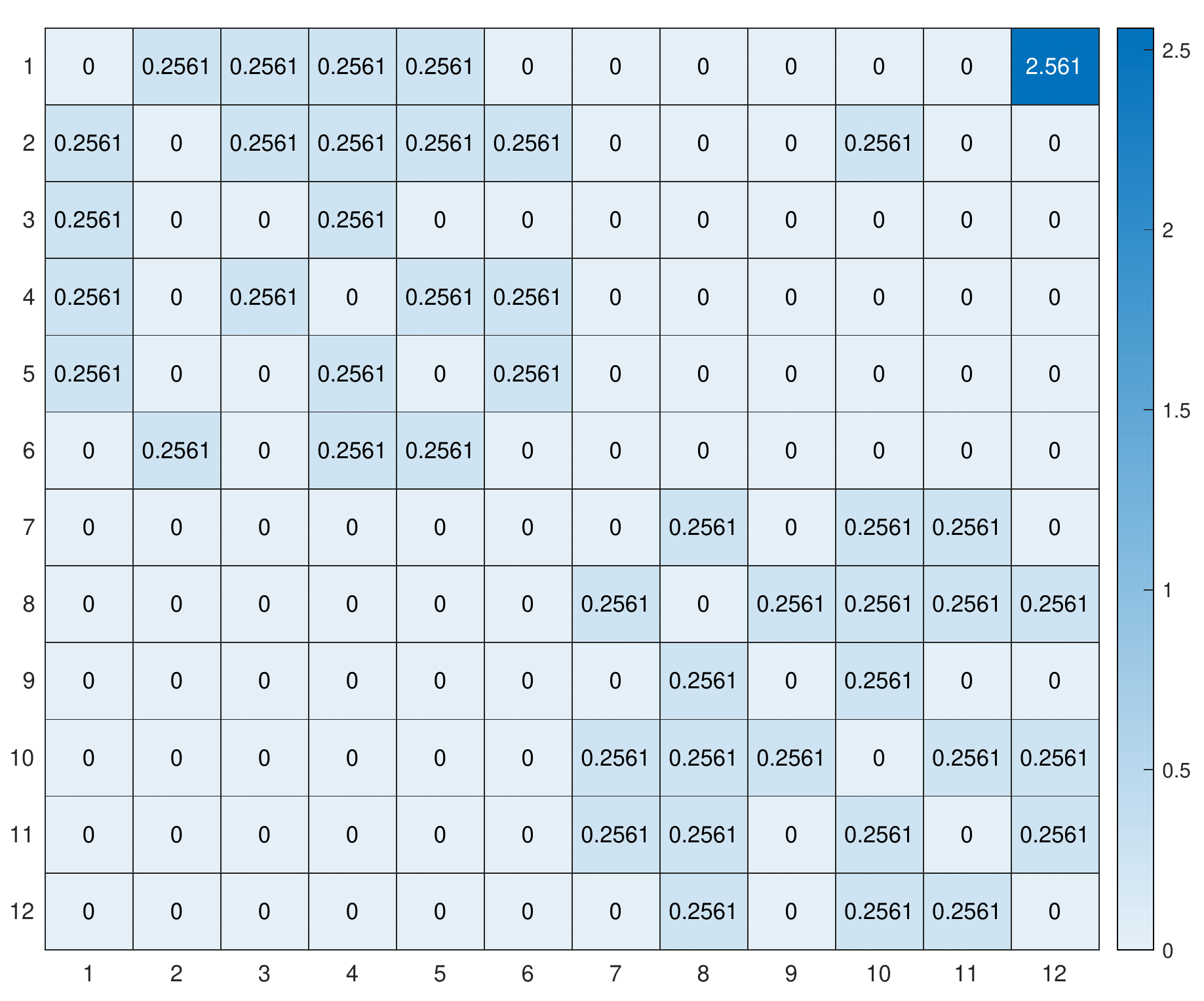}}
\caption{(a) Graph structure of the linear network. (b) Adjacency matrix for the directed weighted graph.}
\label{fig_linear_network_graph}
\end{figure}

\begin{figure}[htp!]
\centering
\includegraphics[scale=.3]{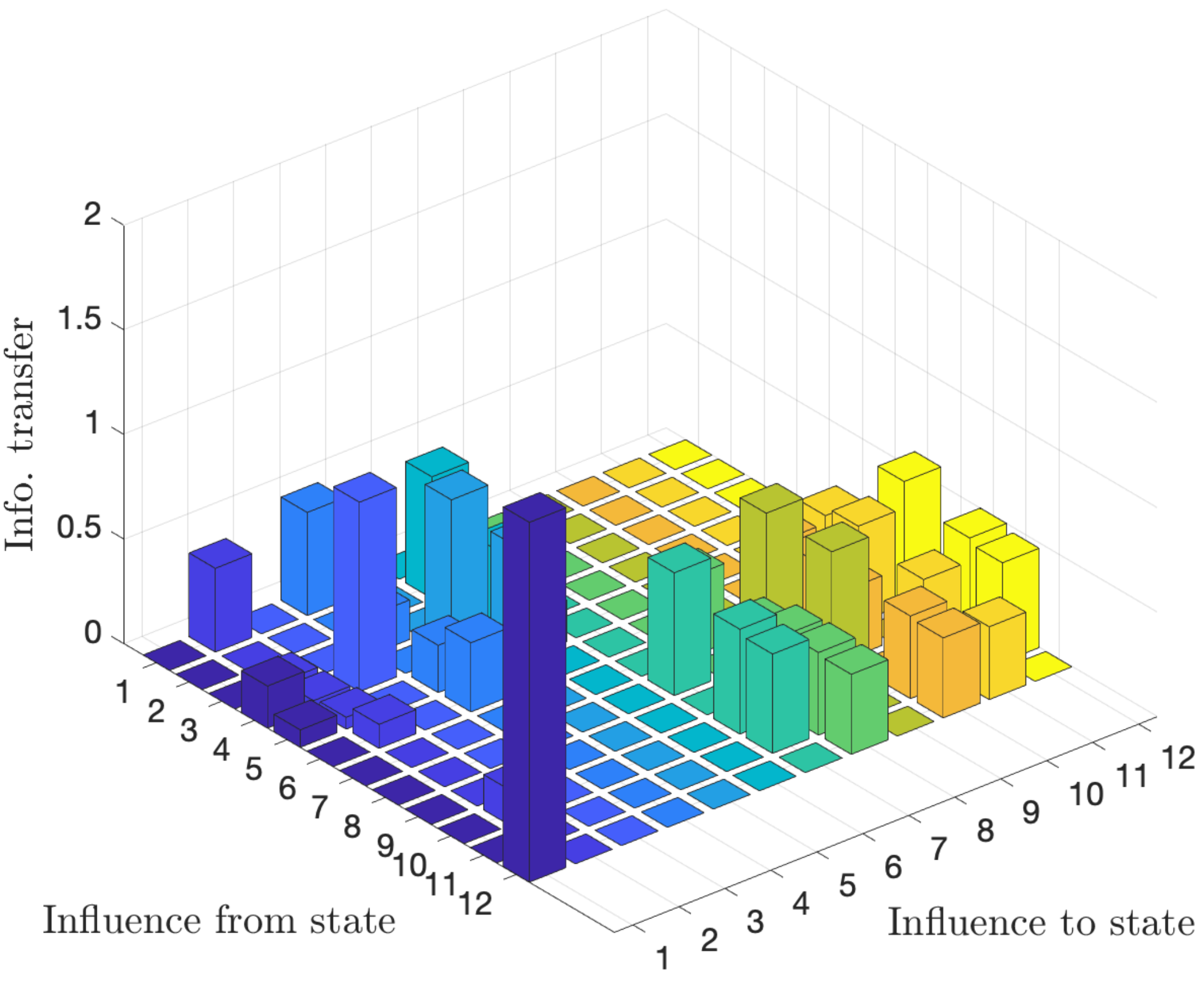}
\caption{Information transfer between the states of the linear dynamical system.}
\label{fig_IT_linear_system}
\end{figure}

We consider a directed weighted adjacency matrix such that the graph structure of the linear oscillator network (\ref{osc_sys}) is as shown in Fig. \ref{fig_linear_network_graph}(a) with the weighted adjacency matrix shown in Fig. \ref{fig_linear_network_graph}(b). 
\begin{figure}[htp!]
\centering
\includegraphics[scale=.275]{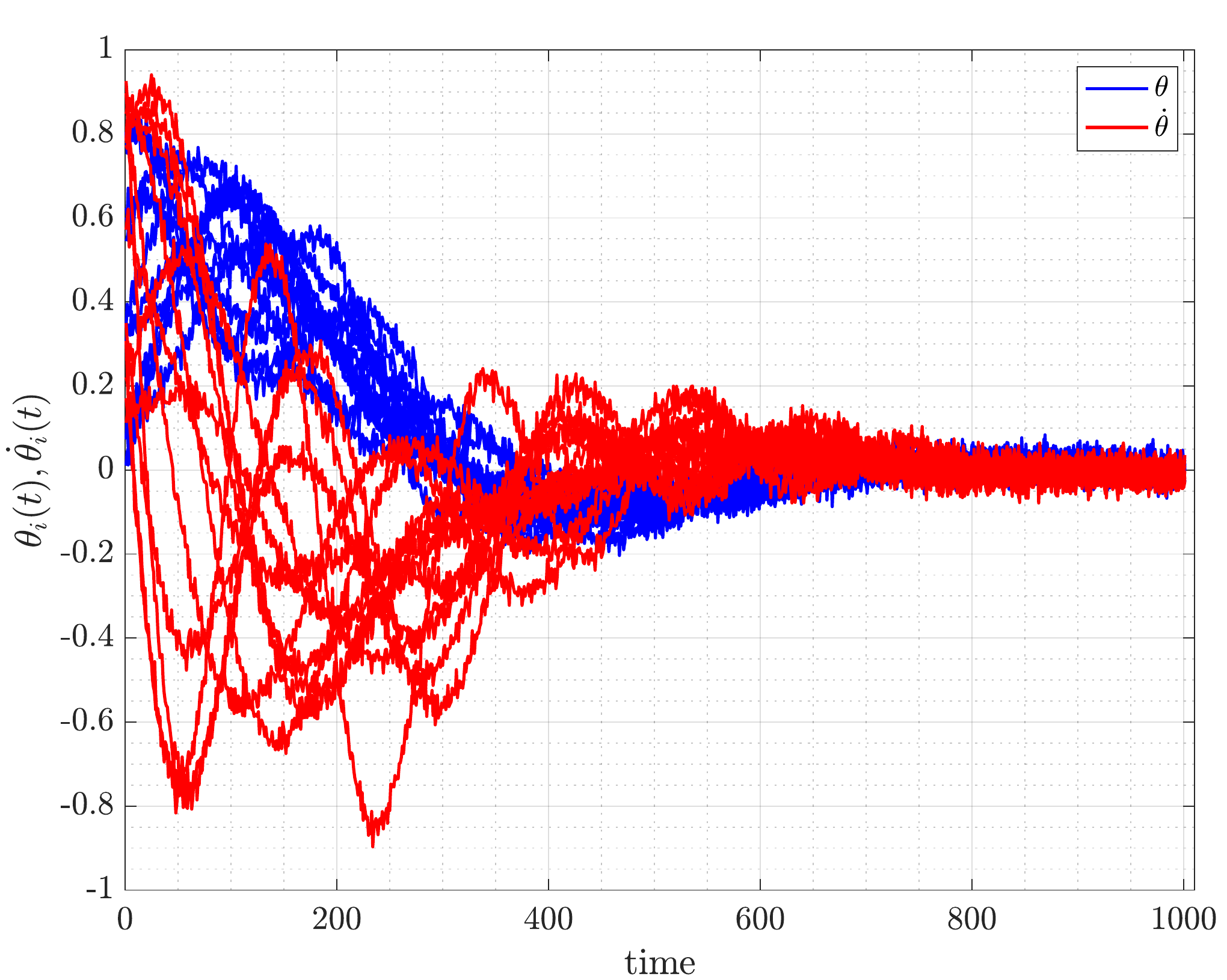}
\caption{Time-series data from the linear dynamical network.}
\label{fig_data_linear}
\end{figure}
For computing the information transfer, a random initial condition was chosen for (\ref{osc_sys}) and data was generated for 1000 time-steps (Fig. \ref{fig_data_linear}) and the information transfers between the oscillators were computed using Algorithm \ref{algo_IT}. The steady state information transfers between the different oscillators in the network is plotted in Fig. \ref{fig_IT_linear_system}. Note that information transfer is not computed between all the 24 states of the network, but between each oscillator, that is, information transfer from $[\theta_i,\dot{\theta}_i]$ to $[\theta_j,\dot{\theta}_j]$ for $i,j = 1, 2, \dots , 12$ and $i\neq j$.

\begin{figure}[htp!]
\centering
\subfigure[]{\includegraphics[scale=.27]{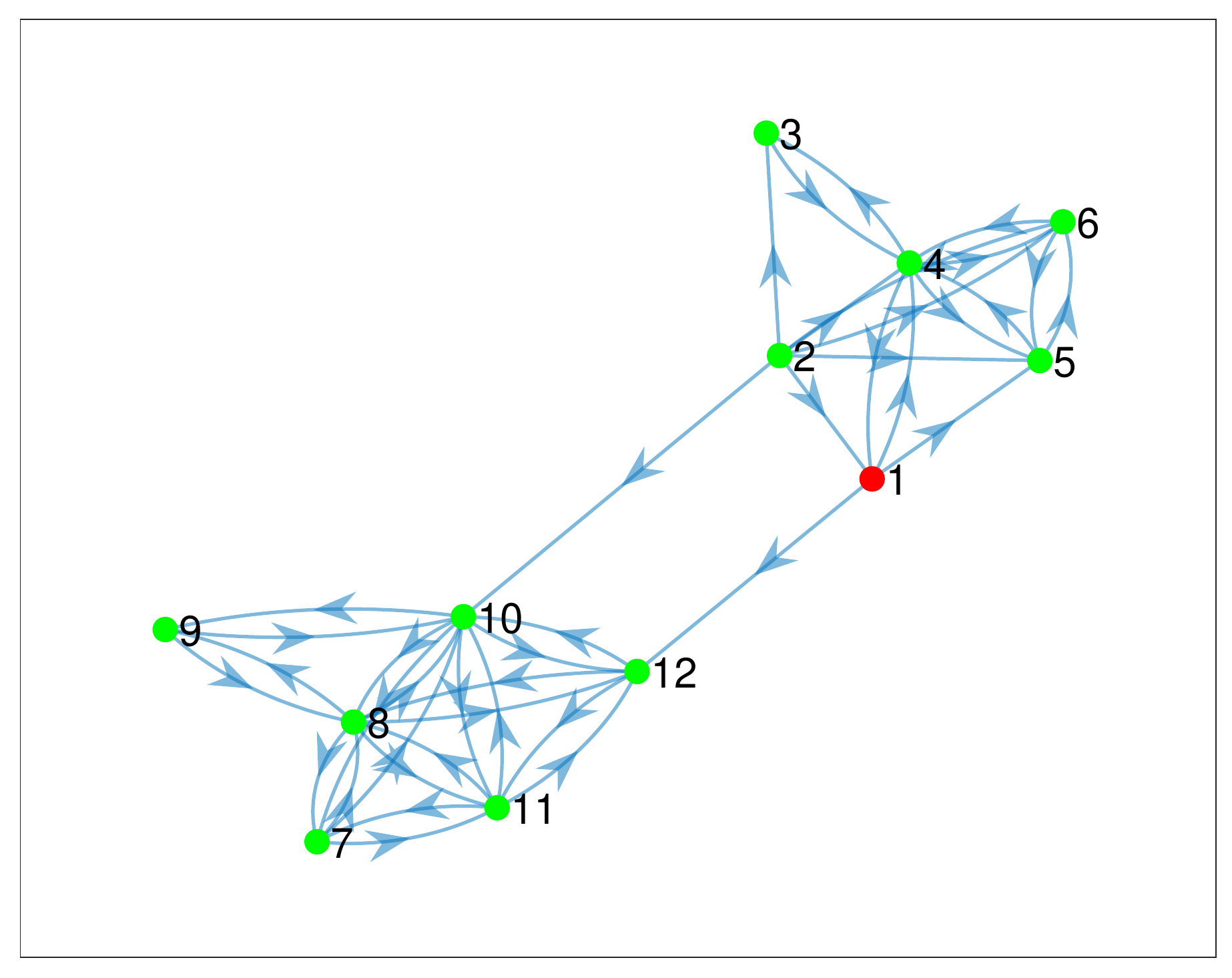}}
\subfigure[]{\includegraphics[scale=.32]{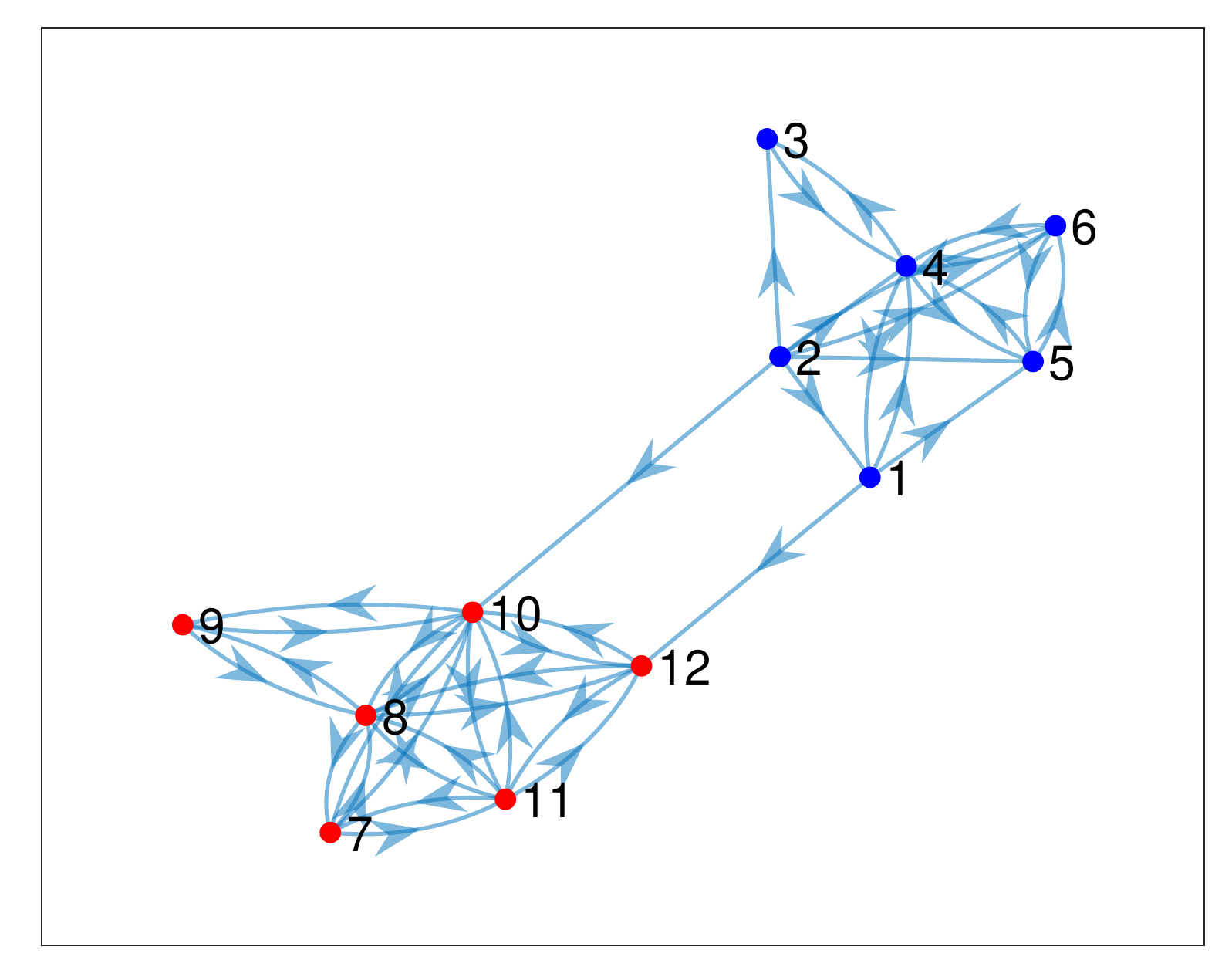}}
\caption{(a) Spectral clustering of the dynamical network based on the adjacency matrix. (b) Spectral clustering of the dynamical network based on influence distance.}
\label{fig_linear_network_cluster}
\end{figure}

With this, we are now in a position to analyze the results of spectral clustering of the oscillator network based on (a) the weighted adjacency matrix and (b) the information distance metric. Firstly, spectral clustering of the adjacency matrix for the dynamical network identified two clusters (Fig. \ref{fig_linear_network_cluster}(a)) with the first oscillator (corresponding to node 1) forming one cluster with a single node and the second cluster consisting of all the other oscillators. However, from the network structure (Fig. \ref{fig_linear_network_graph}(a)), it can be seen that there are two distinct clusters, with each cluster consisting of six nodes (oscillators), but spectral clustering of the adjacency matrix fails to identify these clusters. For influence-based clustering, which considers the dynamical behaviour of the underlying network, the influence distance between each node was calculated with the parameter $\beta = 1$ and spectral clustering of the dynamical network obtained using the influence measure is shown in Fig. \ref{fig_linear_network_cluster}(b). We find that clustering using the influence distance correctly identifies the two clusters with six nodes (oscillators) in each cluster, thus establishing the fact that for clustering of dynamical systems it is imperative to explicitly take into account the dynamical nature of the system.

\subsection{Clustering of Generators for the IEEE 39 Bus System}
In this example, we analyze the cluster structure of the generators of the IEEE 39 bus system. 
\subsubsection{The model and data generation}

The model used in this section is based on the modelling described in \cite{Sauer_pai_book}. The power network is described by a set of differential algebraic equations (DAE) and the power system dynamics is divided into three parts: differential equation model describing the generator and load dynamics, algebraic equations at the stator of the generator and algebraic equations describing the network power flow. We considered a $4^{th}$ order model for the generators with the states of each generator being generator rotor angle $(\delta)$, the angular velocity of the rotor $(\omega)$, the quadrature-axis induced emf $(E_q)$ and the emf of fast acting exciter connected to the generator $(E_{fd})$.  For detailed discussion on the modelling of the power grid we refer the reader to \cite{Sauer_pai_book}.

We also considered IEEE Type-I power system stabilizers (PSS), consisting of a wash-out filter and two phase-lead filters, which are connected to each generator. The input to the  $i^{th}$ PSS controller is $ \omega_i(t)$ (angular speed of the $i^{th}$ generator) and the PSS output $V_{{ref}_i}(t)$ (reference voltage) is fed to the fast acting exciter of the generator. 



\begin{figure}[htp!]
\centering
\includegraphics[scale=.25]{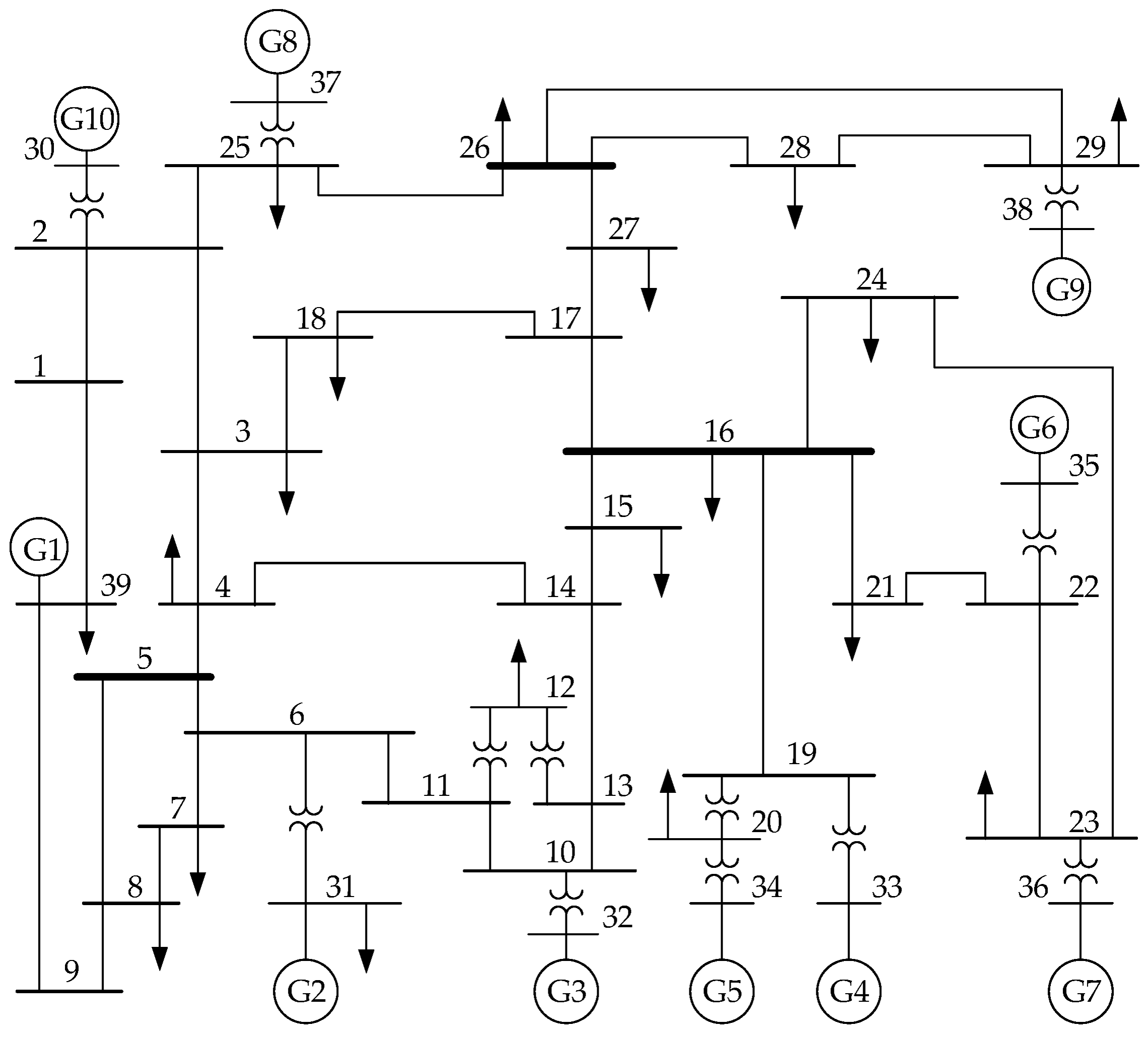}
\caption{IEEE 39 bus system.}\label{39_bus_fig}
\end{figure}

The line diagram of the IEEE 39 bus system is shown in Fig. \ref{39_bus_fig}, which has 10 generators and thus the state space of the system is $\mathbb{R}^{70}$.

For simulation purposes, we consider three different operating conditions of the power network, with different load levels. The first operating point is chosen for $P = 254.1 MW$, where the system is extremely stable, the second operating point is for $P = 900 MW$, where the system is moderately stable and the last operating point is for $P = 1740.68 MW$, where the system is on the verge of becoming unstable. 


For simulation purposes, data was generated at each of the three operating points for 1000 time-steps by solving the set of nonlinear differential equations and steady state information transfer between the ten generators was computed by the procedure outlined in algorithm \ref{algo_IT}.


\subsubsection{k-means clustering}

$k$-means clustering \cite{kmeans} is one of the most commonly used methods for clustering and it aims to divide the nodes into $k$ clusters to minimize the within-cluster distances. In this section, we divide the 39 bus system into three clusters and study how the clusters evolve with changes in the operating condition. As mentioned earlier, we choose three different operating conditions, namely, $P = 254.1 MW$, $900 MW$ and $1740.68 MW$.

\begin{figure}[htp!]
\centering
\subfigure[]{\includegraphics[scale=.35]{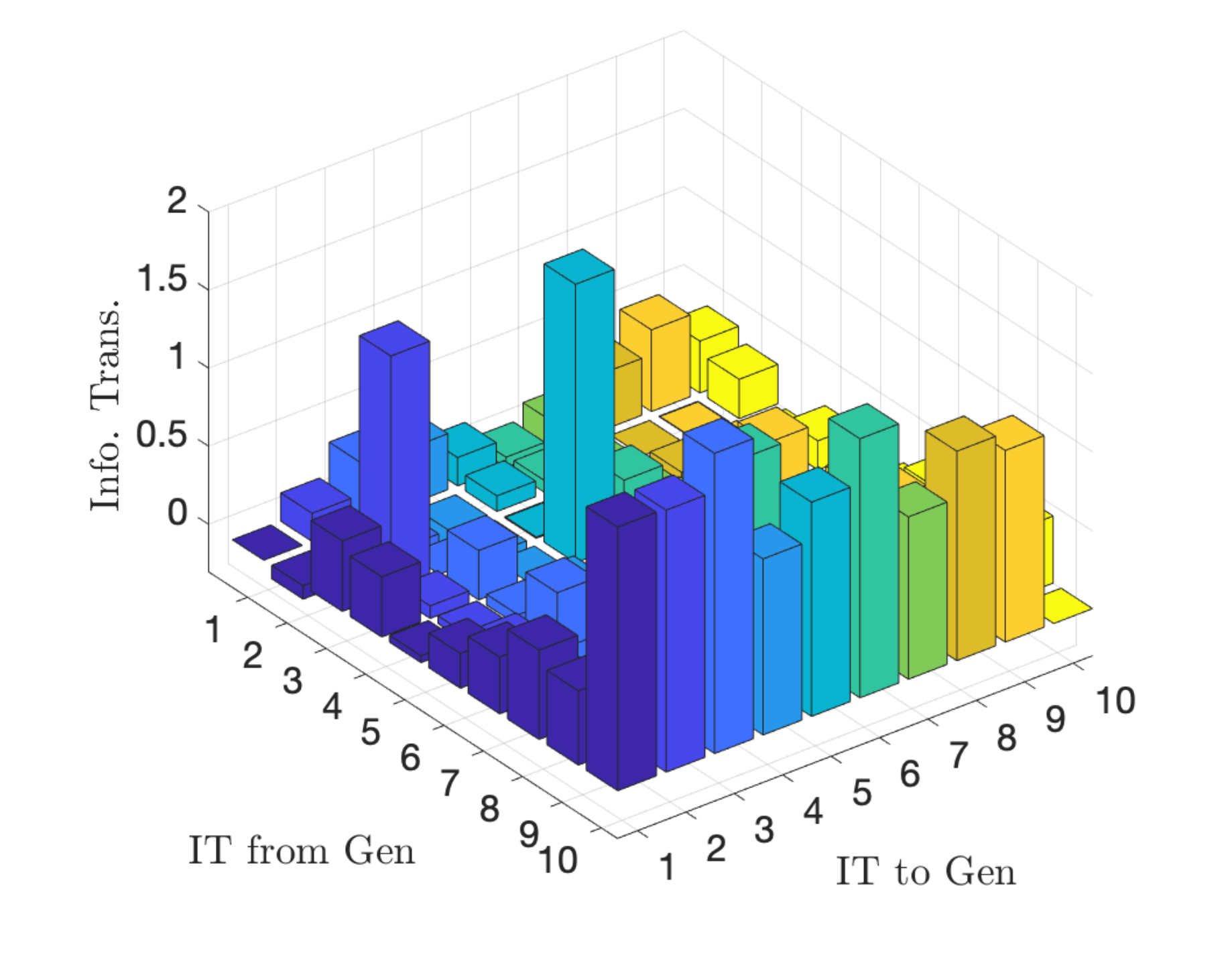}}
\subfigure[]{\includegraphics[scale=.35]{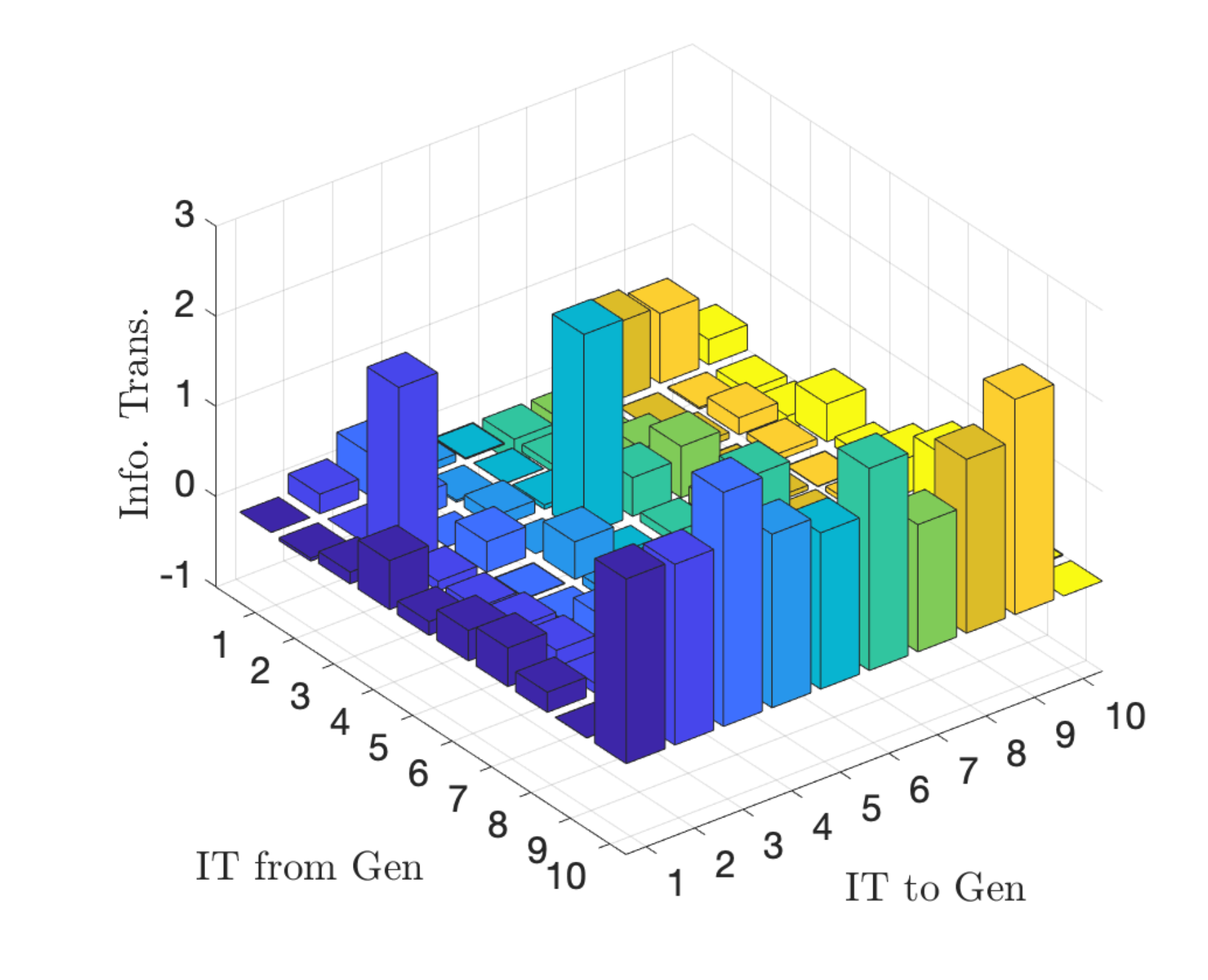}}
\subfigure[]{\includegraphics[scale=.35]{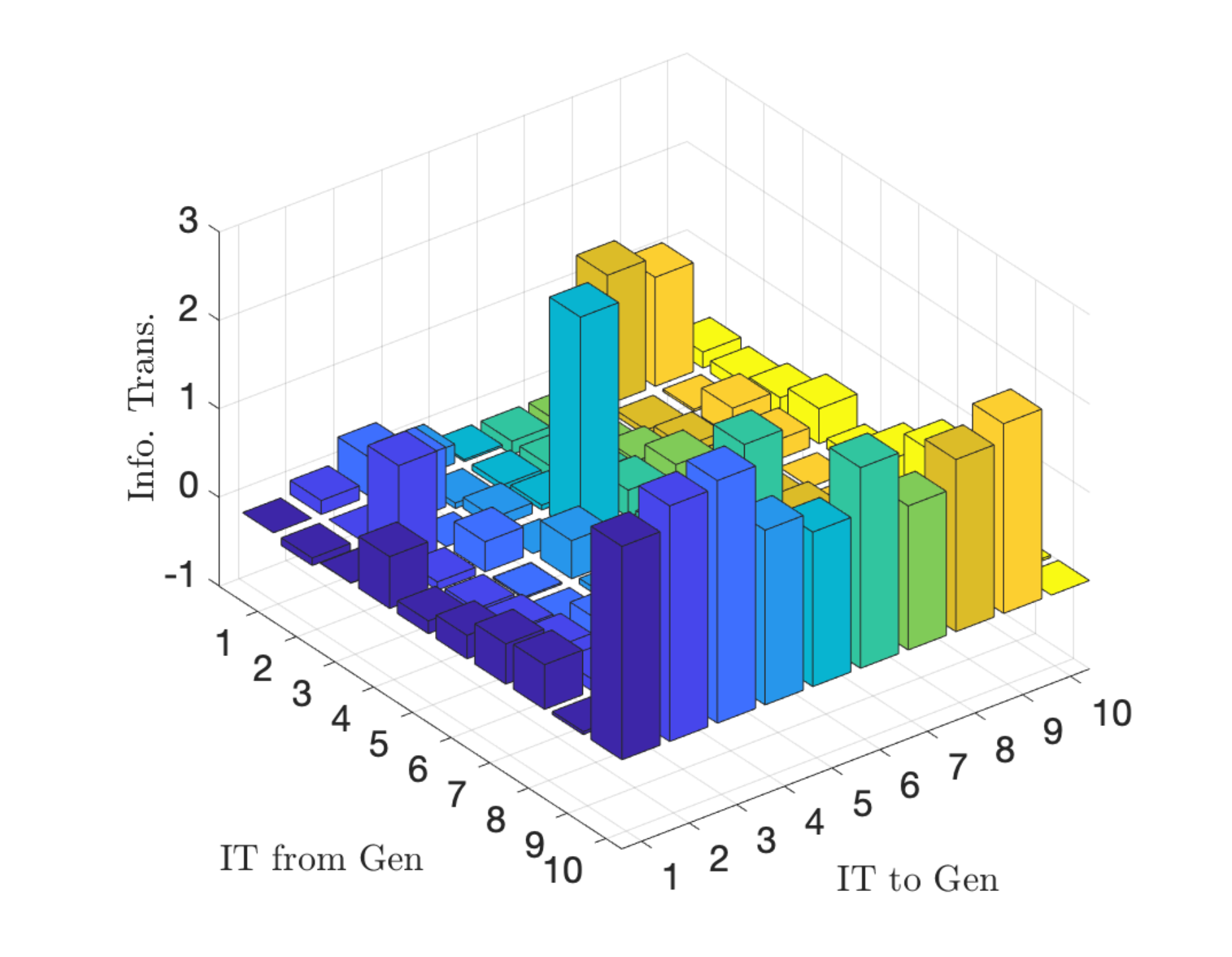}}
\caption{Information transfer between the generator subspaces at (a) $P = 254.1 MW$, (b) $900 MW$ and (c) $1740.68 MW$}\label{fig_IT_op_pt}
\end{figure}

The information transfers between the different generators for the three operating conditions are plotted in Fig. \ref{fig_IT_op_pt}. It can be observed that generator 10 has a large influence on all the other generators over the operating points. Hence, all the generators are \emph{close} to generator 10. As such, generator 10 is the most influential generator.


\begin{figure}[htp!]
\centering
\subfigure[]{\includegraphics[scale=.2]{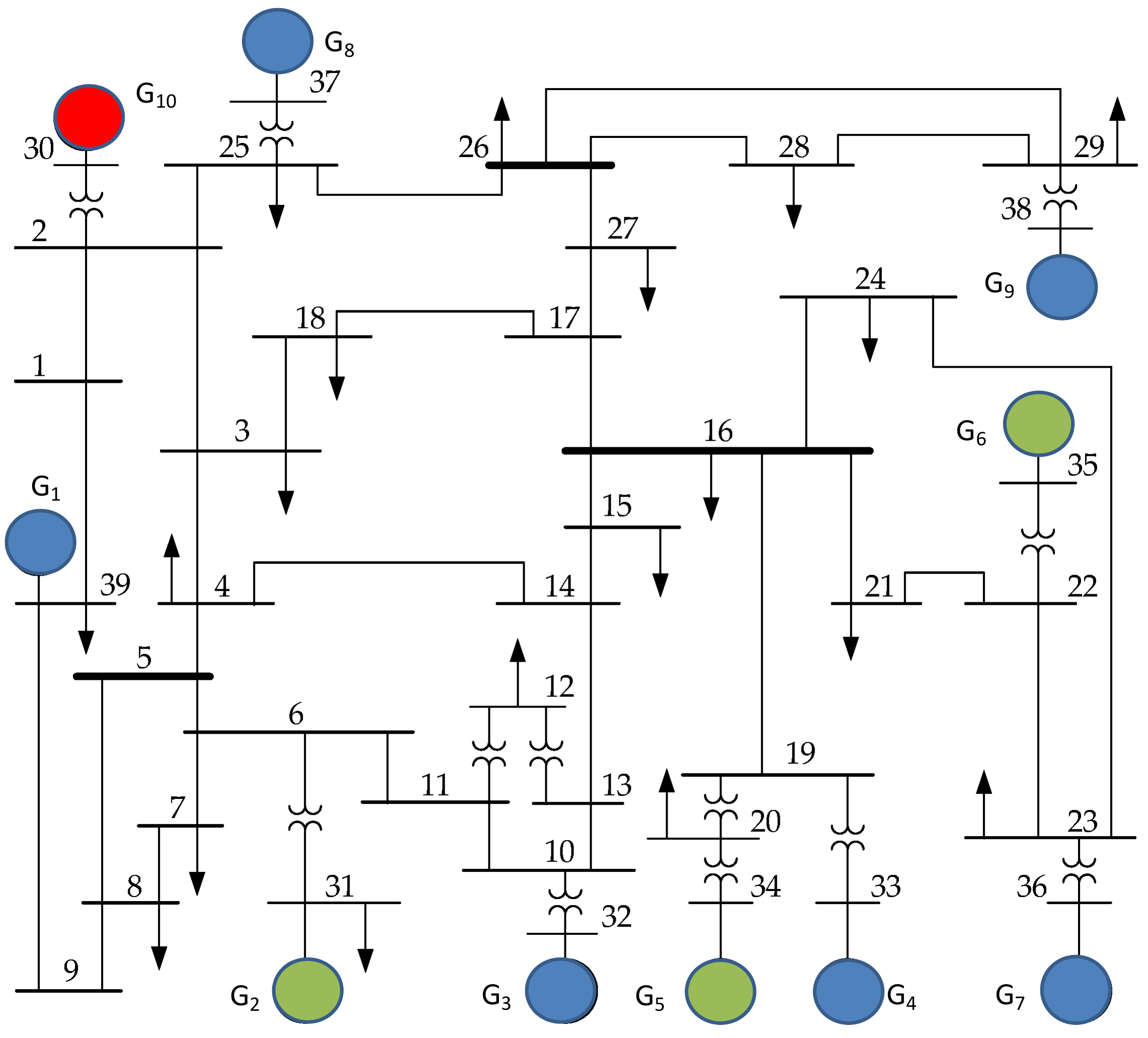}}
\subfigure[]{\includegraphics[scale=.2]{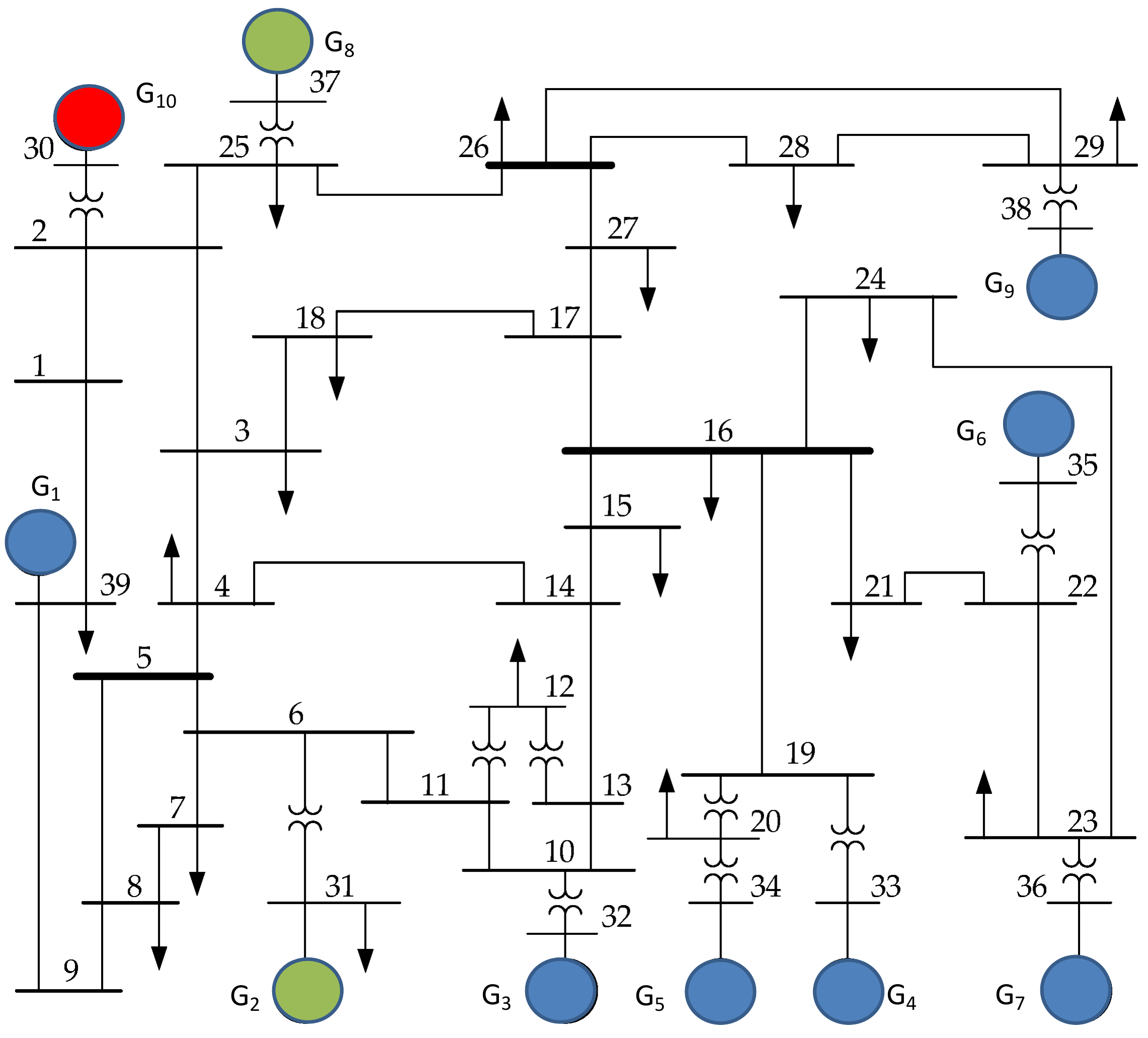}}
\subfigure[]{\includegraphics[scale=.2]{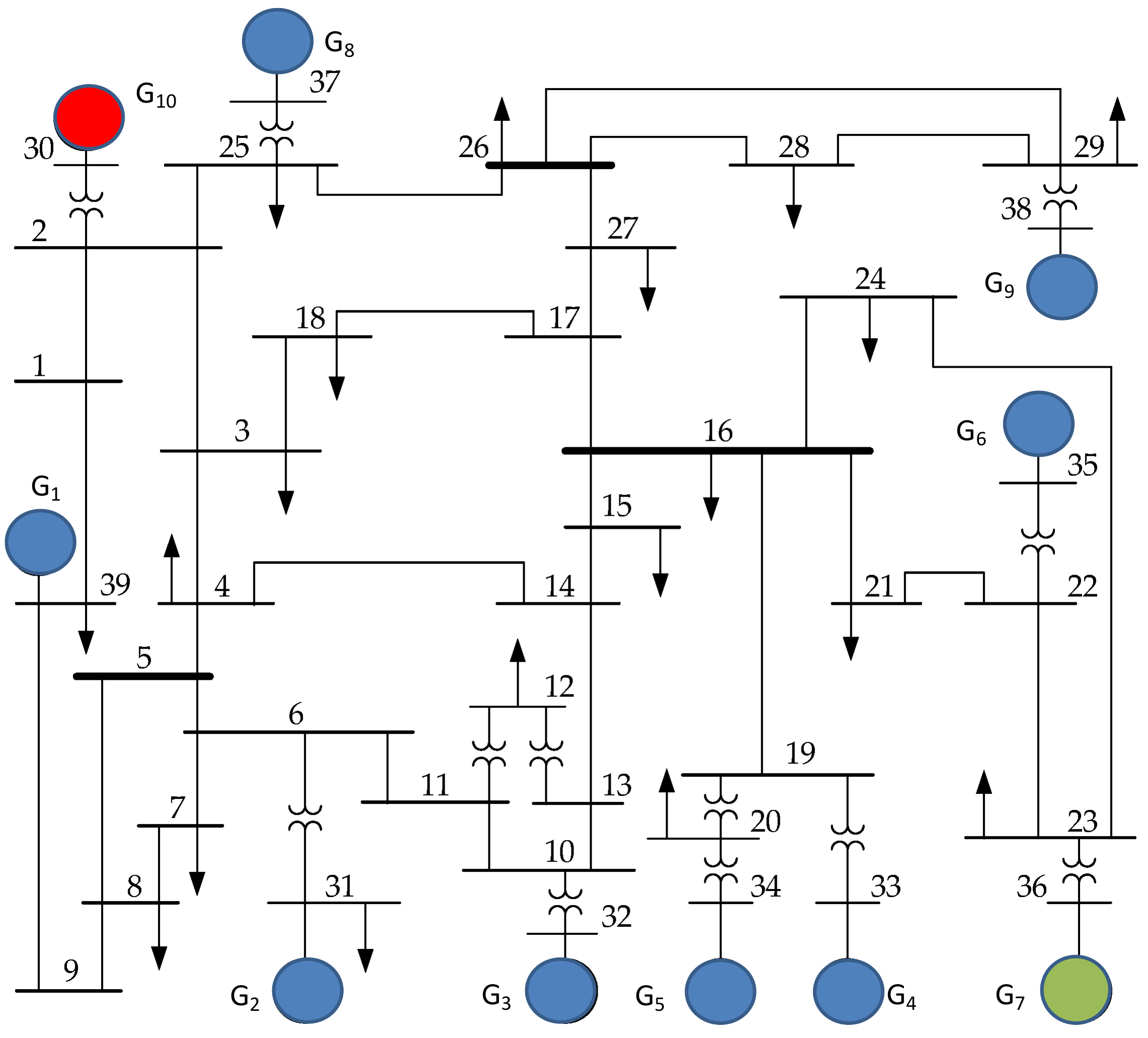}}
\caption{Cluster structure of IEEE 39 bus network for (a) $P = 254.1 MW$, (b) $900 MW$ and (c) $1740.68 MW$}\label{fig_kmeans_op_pt}
\end{figure}

\vspace{.2cm}
Fig. \ref{fig_kmeans_op_pt} shows the clusters at three different operating points, where generators of the same colour belong to the same cluster. It can be observed that generator 10 is always a single cluster and this abides by the intuition that generator 10 influences all the other generators, whereas no other generator has a substantial influence on generator 10. Hence, though all the generators are \emph{close} to generator 10, generator 10 is not close to any other generator. In fact, in \cite{sinha_power_journal_IEEEtran}, it was shown that generator 10 is the most influential generator and is most responsible for the instability of the power network. However, the analysis in \cite{sinha_power_journal_IEEEtran} was model-based, where linearized dynamics was considered. In contrast, in this work, we considered data from the nonlinear model and data-driven information transfer computation and the $k$-means clustering procedure identifies generator 10 as the most influential generator for the IEEE 39 bus power network. It can be further observed that as the load increases the size of the cluster changes. When the system is operating at a very stable region, the two clusters (apart from generator 10) are of similar sizes, but as the load increases the size changes and at the verge of instability $(P = 1740.68 MW)$ there are two clusters with a single generator, namely, generator 10 and generator 7, and all the other 8 generators belong to one cluster. This is because at this operating condition, though only generator 10 has any substantial influence on generator 7, the influence of generator 10, among all the generators, is the lowest on generator 7. Furthermore, at this operating point, generator 7 is not transferring much information to any of the other generators. Thus neither generator 7 is \emph{substantially close} to any other generator nor any other generator is \emph{substantially close} to generator 7 and hence generator 7 forms a separate cluster with itself as the only member of the cluster.



\subsubsection{Hierarchical clustering}

Though $k$-means clustering is popular, this procedure suffers from some deficiencies as far as overall control of the clustering process is concerned. Firstly, the number of clusters has to be decided beforehand. Secondly, $k$-means clustering may not completely encode the cluster structure of the underlying network and instead output a single partition of the network \cite{sanchez2014hierarchical}. Moreover, to reveal the finer structure of the network at different levels of resolution, we use hierarchical clustering \cite{hierarchical} to cluster the generators of the IEEE 39 bus network. 

As before, we analyze the hierarchical structure of the IEEE 39 bus network for the three operating conditions.

\begin{figure}[htp!]
\centering
\subfigure[]{\includegraphics[scale=.225]{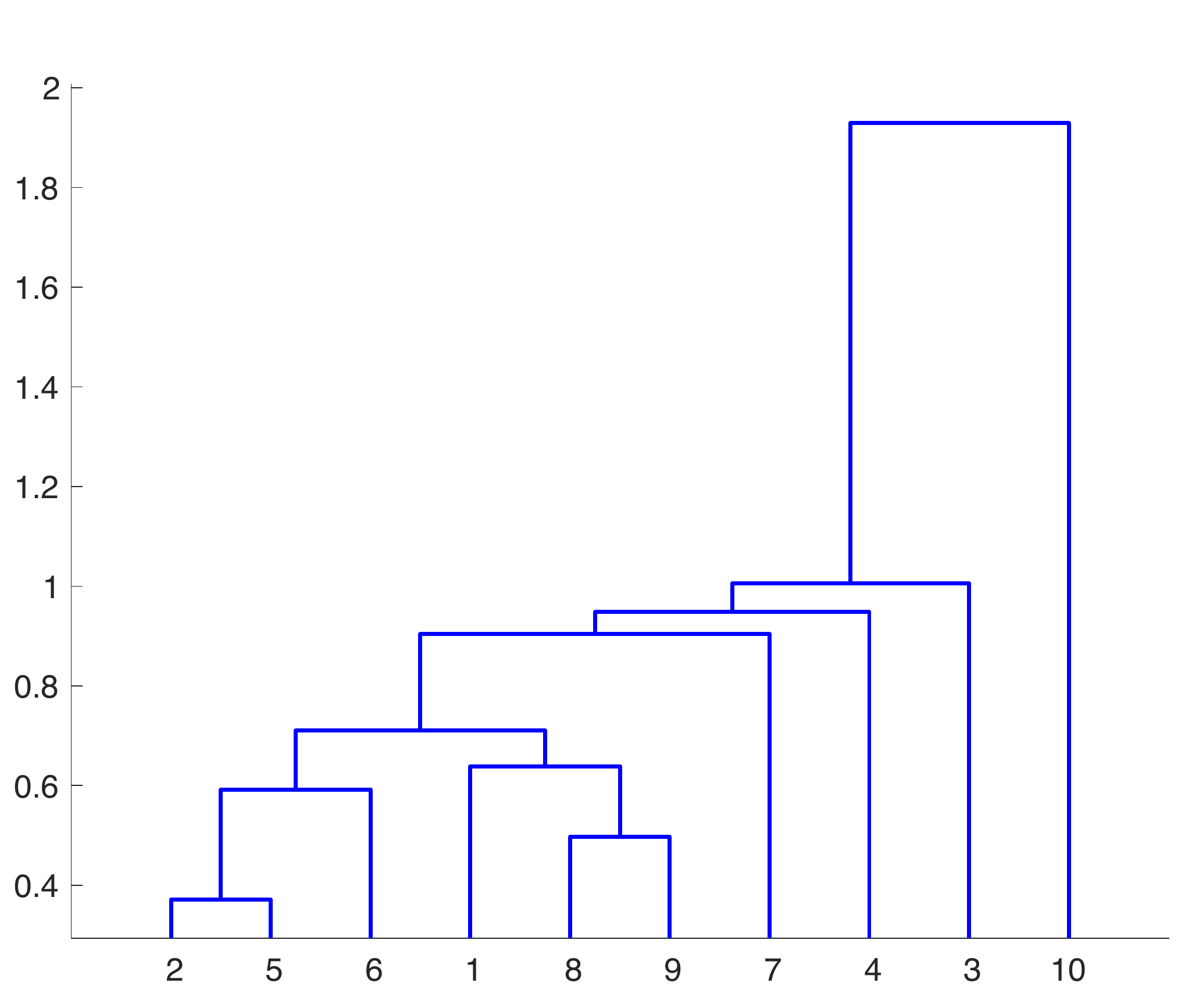}}
\subfigure[]{\includegraphics[scale=.225]{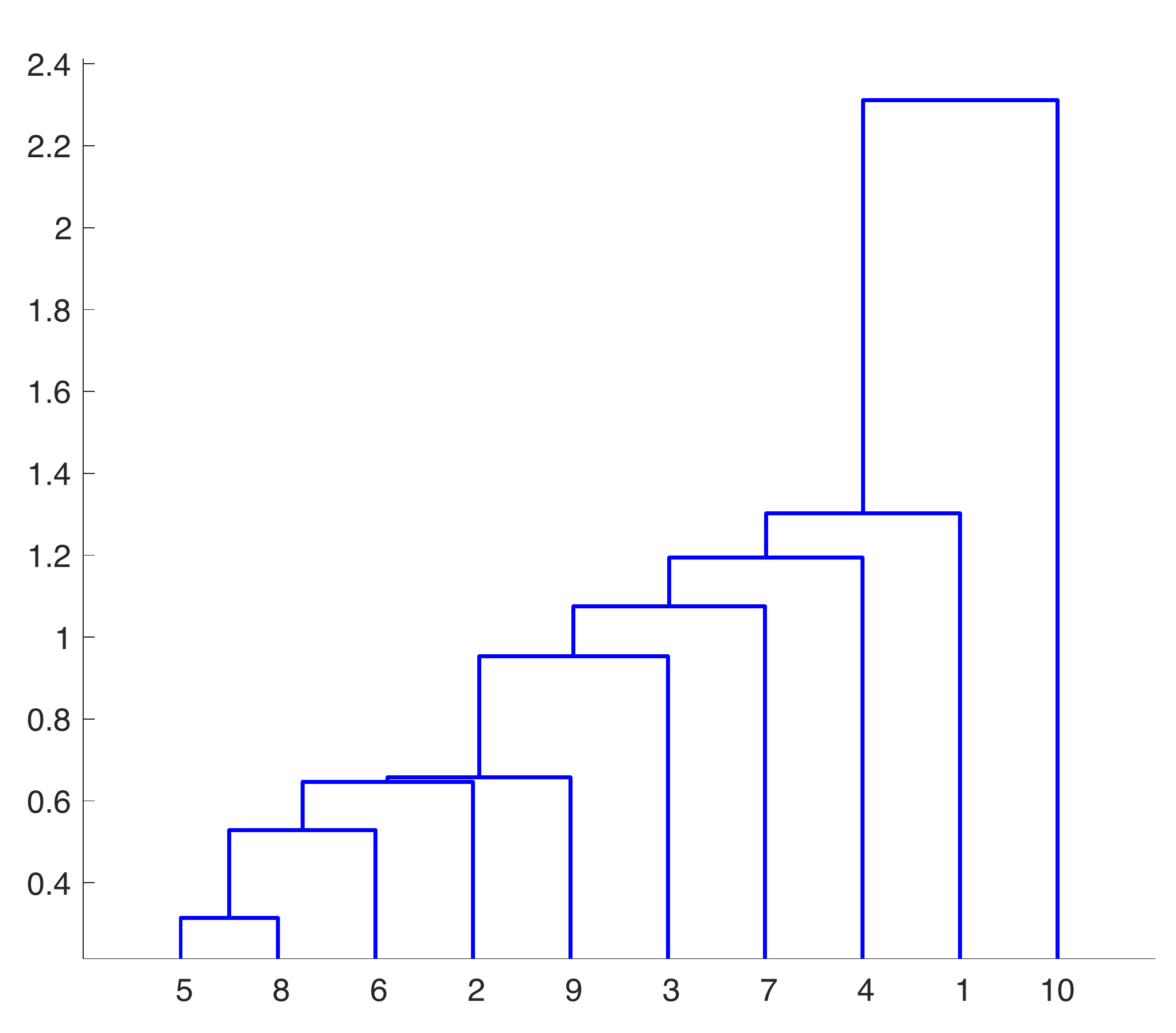}}
\subfigure[]{\includegraphics[scale=.225]{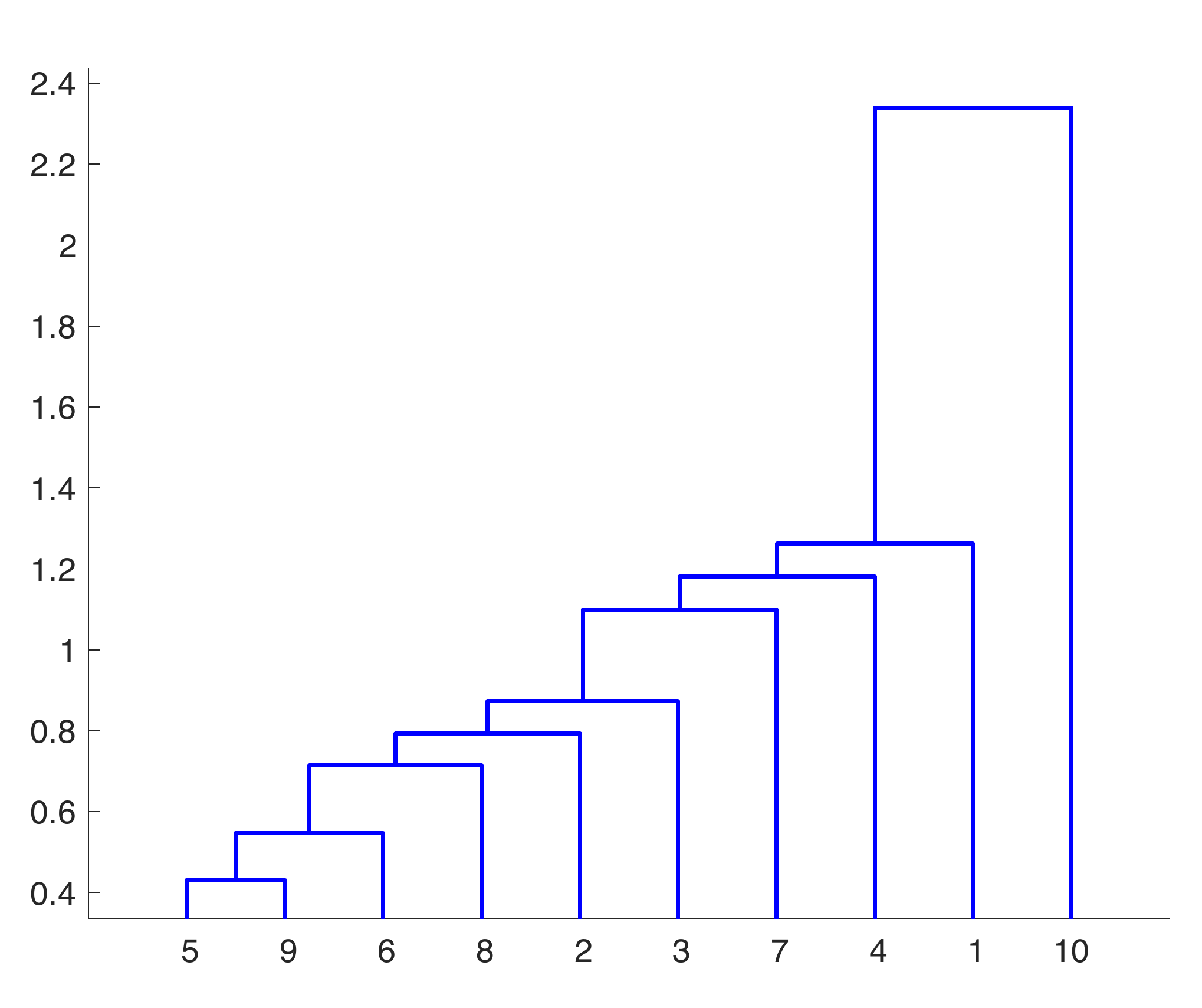}}
\caption{Hierarchical clustering of IEEE 39 bus network for (a) $P = 254.1 MW$, (b) $900 MW$ and (c) $1740.68 MW$}\label{fig_hierarchical_op_pt}
\end{figure}

The hierarchical structures of the network, for different operating conditions, are represented in the form of \emph{dendrograms}, as shown in Fig. \ref{fig_hierarchical_op_pt}. A dendrogram is a tree structure where the bottom leaves represent the individual nodes (generators). These are clusters of size one and at each upper level, \emph{closest} clusters are merged together to get the hierarchical structure. From Fig. \ref{fig_hierarchical_op_pt}(a)-(c), it can be seen that over all the operating points generator 10 is at the highest level of the hierarchical structure. This is concurrent with the fact that in the IEEE 39 bus system, generator 10 is the most influential generator and influences all the other generators. However, the clusters at lower levels change with the operating condition. The changing clusters show that even though the underlying topology of the network remains the same throughout the operating conditions, the dynamical nature of the network changes with the operating points. Moreover, the hierarchical structure identifies the generators that can be used for local control. For example, if the network is operating at $P = 900 MW$, and one needs to control generators 5 and 8 by using a single controller, it is reasonable to control them from generator 6 (see Fig. \ref{fig_hierarchical_op_pt}(b)). This is because generator 6 lies directly above generators 5 and 8 in the dendrogram and thus affects them directly and moreover, implementing a control at generator 6 will not affect the other generators (apart from generators 5 and 8) since it has a small influence on the other generators. This is because generator 6 lies just above generators 5 and 8 and all the other generators are above generator 6 in the dendrogram plot of Fig. \ref{fig_hierarchical_op_pt}(b). 

Further, we note that for $P = 1740.68 MW$, when the system is on the verge of being unstable, the hierarchical structure is almost a tree (Fig. \ref{fig_hierarchical_op_pt}(c)), except for generators 5 and 9, which lie on the same level and form a single cluster. Generator 10 is the cause of instability \cite{sinha_power_journal_IEEEtran} and hence is at the top of the hierarchical tree (Fig. \ref{fig_hierarchical_op_pt}(c)) and the hierarchical tree structure shows spreading of the instability through the network in a cascading effect.

\subsection{Clustering of Features in WRF-Chem Model for Amazon Rain Forest}
In this subsection, we present preliminary results on the clustering of dynamic variables found in the atmosphere over the Amazon rain forest. Data were obtained from the Weather Research and Forecasting Model coupled to chemistry (WRF-Chem), which is a community three-dimensional chemical transport model that couples clouds, gas-phase and particle-phase chemistry, meteorology and radiation online and interactively \cite{grell2005fully}. The model is used to study coupled physical and chemical processes
such as aerosol-cloud interactions, clouds and convection.

{\small
\begin{table}[h!]
\centering
\caption{Dynamic Variables Considered}\label{variable_acronym}
\begin{tabular}{|l|l|} 
\hline
Full Name & Acronym\\
\hline 
temperature &  tk \\ 
relative humidity &  rh \\ 
ambient pressure &  p \\  
isoprene epoxydiol (IEPOX) gas &  iepoxgas \\ 
2-methyltetrol gas &  tetrolgas \\ 
glass transition temperature &  TGLASSCOAT \\ 
of organic aerosol & \\
organic aerosol &  TOTOAtotal \\ 
particle water  &  watertotal \\ 
particle sulfate &  so4total \\ 
particle nitrate &  no3total \\ 
particle ammonium & nh4total  \\ 

particle IEPOX organosulfate & iepoxostotal  \\
particle tetrol oligomer &  tanvtotal \\ 
particle tetrol &  tetroltotal\\
\hline
\end{tabular}
\end{table}
}

The WRF-Chem model was run at a moderately high resolution of 10 km grid spacing encompassing a $1500\times 1000$ km domain from near-surface to the free troposphere (altitudes of ~15 km) over the Amazon rainforest to simulate the formation of secondary organic aerosol (SOA) \cite{shrivastava2017recent}. The vertical altitude range from 0-15 km was divided into 44 vertical levels, with half the number of vertical levels placed in the lowest 2 km altitude. Data for all the variables shown in Table \ref{variable_acronym}, were selected for seven consecutive days and in this set of simulations we considered five different computational altitude plains, namely, levels $0,11,22,33,44$, where 0 is near the surface, and 44 is at 15 km altitude. The data-set had 2832192 data points for each level and for computation of the Koopman operator we normalized the data. 

\begin{figure}[htp!]
\centering
\subfigure[]{\includegraphics[scale=.2]{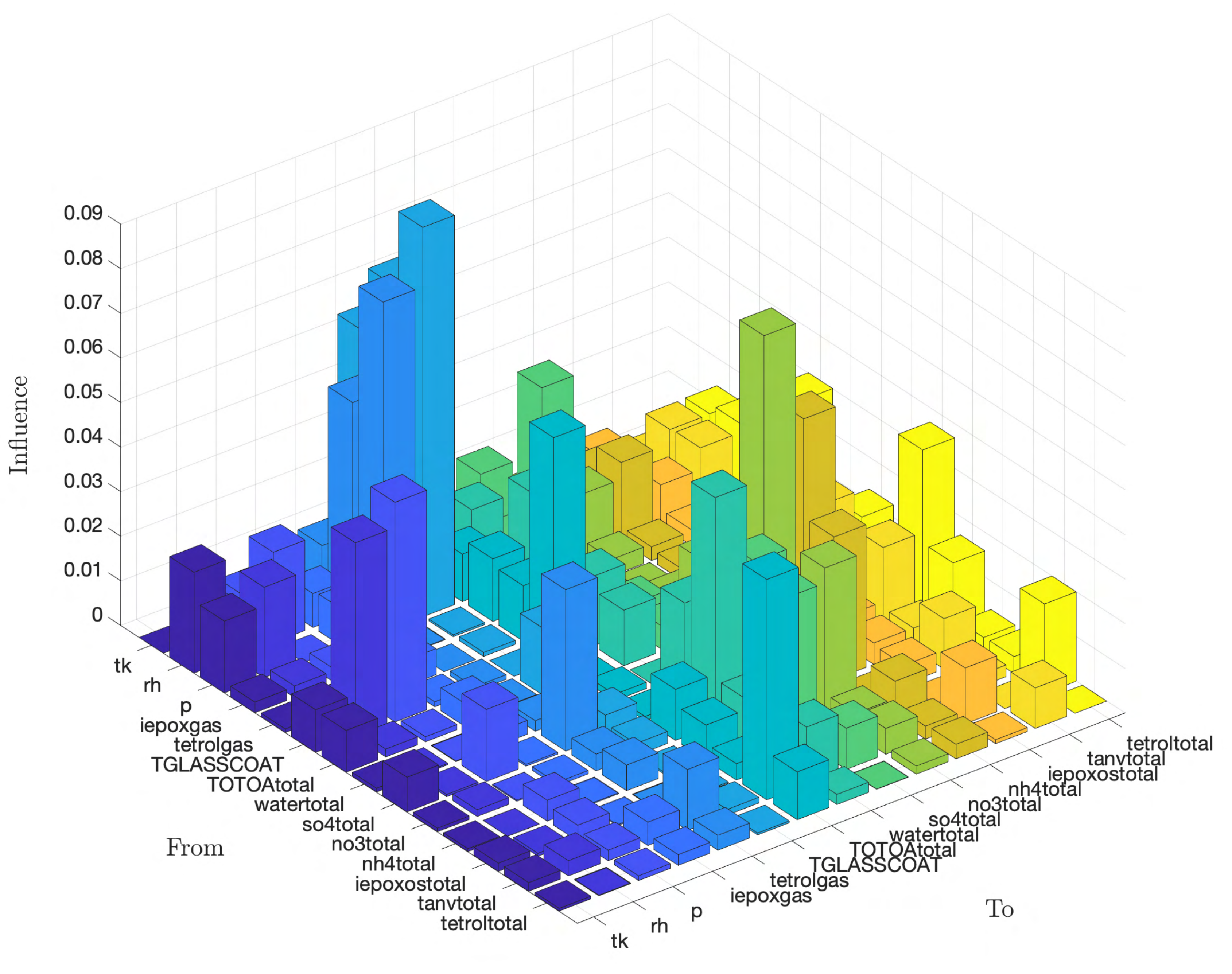}}
\subfigure[]{\includegraphics[scale=.27]{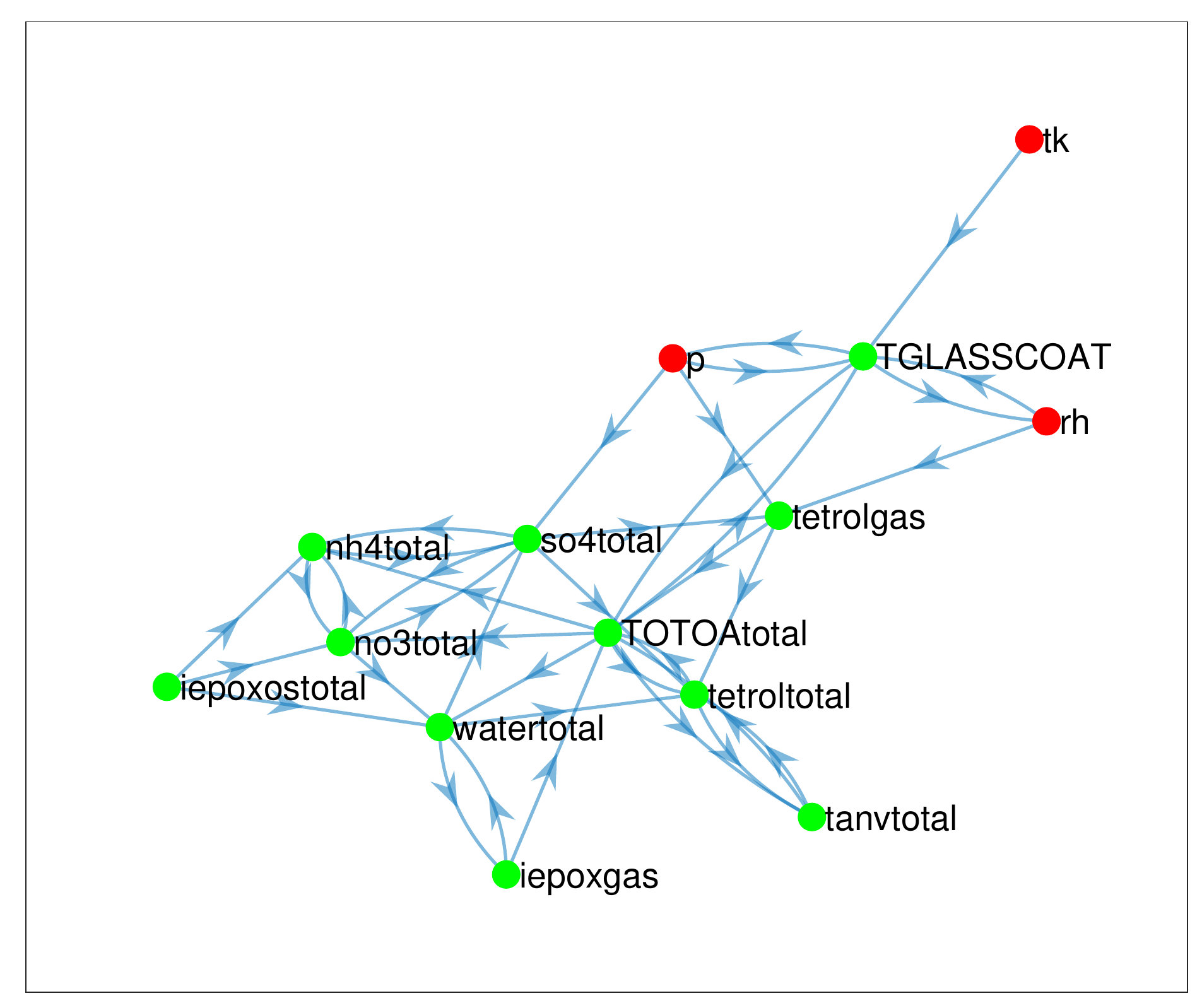}}
\caption{(a) Information transfer between the various dynamic features of the WRF-Chem model. (b) Influence graph and spectral clustering of the dynamic variables.}\label{fig_IT_climate}
\end{figure}

With this, the information transfer between the various features is shown in Fig. \ref{fig_IT_climate}(a) and the influence graph and the clustering of the variables, based on the information distance measure is shown in Fig. \ref{fig_IT_climate}(b). To obtain the influence distance measure, we set the information transfer values which were less than $0.01$ to be equal to zero and chose the parameter $\beta$ to be unity. Furthermore, we used spectral clustering on the weighted directed influence distance graph to cluster the variables. In this preliminary analysis, we find that the proposed method groups the temperature, pressure and relative humidity into one cluster and groups all the chemical species into a separate cluster. This makes sense because among all the variables considered, temperature, pressure and relative humidity are the only \emph{large scale} variables and hence it is natural for them to belong to the same cluster. The other variables are chemical species and hence are quite different in nature to temperature, pressure and relative humidity and hence they all belong to a separate cluster.

\section{Conclusions}\label{section_Conclusions}
In this paper, we propose a novel data-driven approach for clustering dynamical systems, which takes into account the dynamics in the learning optimal clusters. To that end, we utilize tools from Koopman operator framework to learn the underlying (possibly nonlinear) dynamical system from time-series data and leverage that information in defining a weighted graph for the system, where the weights capture the influence between the states. We establish the efficacy of the proposed method on a network of linear oscillators, where we show that the proposed approach correctly identifies the clusters in the network (as opposed to clustering using the adjacency matrix of the network). We also analyzed the non-linear IEEE 39 bus system, where we used both $k$-means and hierarchical clustering algorithms to determine the community structure of the generators of the IEEE 39 bus network. However, in a power network data is obtained from Phasor Measurement Units (PMUs) and in the future, we propose to use real PMU data to cluster the buses of a power grid and study how the clusters can be used for the design and implementation of local control actions. Furthermore, we also presented a preliminary analysis of data over the Amazon rain forest and showed that the proposed method divides the dynamic variables into two subgroups so that one cluster contains the large-scale variables and the other cluster contains the chemical species.

\bibliographystyle{IEEEtran}
\bibliography{ref,ref1,subhrajit_ref,subhrajit_power2}

\end{document}